\documentclass[twocolumn]{aastex62}

\usepackage{wrapfig} 
\usepackage{savesym}
\savesymbol{tablenum} %multicolumn ruined everything ...?
\usepackage{siunitx}
\usepackage{multirow}
\usepackage[caption=false]{subfig}
\usepackage{scrextend}
\usepackage{hyperref}
\graphicspath{{./}{figures/}}

\accepted{July 21, 2020}
\submitjournal{ApJ}

\shorttitle{Sample article}
\shortauthors{Drake et al.}
\def\BR{\mbox{BR$1202-0725$}}
\def\haloflux{F$_{\rm{Ly\alpha}} = $ $1.39\,\pm$ $\num{0.01e-15}$}
\def\halofluxno{$13.9\pm0.18$} % in 10^-16 for table

\def\LAEa{LAE\,1}
\def\LAEaEW{EW$_0$(Ly$\alpha$)$ = 72.9\pm2.2$\,\AA}
\def\LAEaEWno{$72.9\pm2.2$}

\def\LAEb{LAE\,2}

\def\LAEc{LAE\,3}
\def\LAEcEW{EW$_0$(Ly$\alpha_{\,5\sigma}^{\,lim})\,\geq\,34.05$\,\AA}
\def\LAEcEWno{$\geq 34.05$}

\def\LAEd{SRC\,6}

\def\fluxunits{erg\,s$^{-1}$\,cm$^{-2}$}
\def\lumunits{erg\,s$^{-1}$}
\def\SBunits{erg\,s$^{-1}$\,cm$^{-2}$\,arcsec$^{-2}$}
\def\kms{km\,s$^{-1}$}
\def\SFRunits{M${_\odot}$\,yr$^{-1}$}

\def\CII{[C{\sc{ii}]}}
\def\CIV{C{\sc{iv}}}
\def\HeII{He{\sc{ii}}}
\def\HII{H{\sc{ii}}}

\def\NII{[N{\sc{ii}]}}

\def\MUSE{{\sc{muse}}}
\def\ALMA{{\sc{alma}}}
\def\HST{{\sc{hst775w}}}
\def\HSTb{{\sc{hst814w}}}
\begin{document}

\title{The Ionised- and Cool-Gas Content of The BR1202-0725 System \\ as seen by MUSE and ALMA.}

\correspondingauthor{Alyssa Drake}
\email{drake@mpia.de}

\author[0000-0002-0174-3362]{A. B. Drake}
\affil{Max Planck Institute f{\"u}r Astronomie, K{\"o}nigstuhl, Heidelberg, Germany}
\author[0000-0003-4793-7880]{F. Walter}
\affil{Max Planck Institute f{\"u}r Astronomie, K{\"o}nigstuhl, Heidelberg, Germany}
\author[0000-0001-8695-825X]{M. Novak}
\affil{Max Planck Institute f{\"u}r Astronomie, K{\"o}nigstuhl, Heidelberg, Germany}
\author[0000-0002-6822-2254]{E. P. Farina}
\affil{Max Planck Institute for Astrophysics, Karl-Schwarzschild-Str, Garching, Germany}
\affil{Max Planck Institute f{\"u}r Astronomie, K{\"o}nigstuhl, Heidelberg, Germany}
\author[0000-0002-9839-8191]{M. Neeleman}
\affil{Max Planck Institute f{\"u}r Astronomie, K{\"o}nigstuhl, Heidelberg, Germany}
\author[0000-0001-9585-1462]{D. Riechers}
\affil{Department of Astronomy, Cornell University, Space Sciences Building, Ithaca, NY 14853, USA}
\affil{Max Planck Institute f{\"u}r Astronomie, K{\"o}nigstuhl, Heidelberg, Germany}
\author[0000-0001-6647-3861]{C. Carilli}
\affil{National Radio Astronomy Observatory, Pete V. Domenici Array Science Center, P.O. Box O, Socorro, NM 87801, USA}
\affil{Cavendish Astrophysics Group, University of Cambridge, Cambridge, CB3 0HE, UK}
\author[0000-0002-2662-8803]{R. Decarli}
\affil{INAF - Osservatorio Astronomico di Bologna, Via Piero Gobetti, 93/3, 40129 Bologna BO, Italy}
\author[0000-0002-5941-5214]{C. Mazzucchelli}
\affil{European Southern Observatory, Alonso de Cordova 3107, Vitacura, Region Metropolitana, Chile}
\author[0000-0003-2984-6803]{M. Onoue}
\affil{Max Planck Institute f{\"u}r Astronomie, K{\"o}nigstuhl, Heidelberg, Germany}

\begin{abstract}

We present \MUSE\ observations of the gas-rich major-merger \BR\ at $z\sim4.7$, which constitutes one of the most overdense fields known in the early Universe. We utilise these data in conjunction with existing \ALMA\ observations to compare and contrast the spatially resolved ionised- and cool-gas content of this system which hosts a quasar (QSO), a sub-millimeter galaxy (SMG), the two known optical companions (``\LAEa", ``\LAEb"), and an additional companion discovered in this work ``\LAEc" just 5\,\arcsec\ to the North of the QSO. We find that QSO \BR\ exhibits a large Ly$\alpha$ halo, covering $\approx55$ pkpc on-sky at surface brightness levels of SB$\geq$\num{1E-17}\SBunits. In contrast, the SMG, of similar far-infrared luminosity and star formation rate (SFR), does not exhibit such a Ly$\alpha$ halo. The QSO's halo exhibits high velocity widths ($\sim1000$ \kms) but the gas motion is to some extent kinematically coupled with the previously observed \CII\ bridge between the QSO and the SMG. We note that the object known in the literature as \LAEb\ shows no local peak of Ly$\alpha$ emission, rather, its profile is more consistent with being part of the QSO's extended Ly$\alpha$ halo. The properties of \LAEc\ are typical of high-redshift LAEs; we measure \mbox{F$_{\rm{Ly\alpha}}$(\LAEc) = $0.24\pm$\num{0.03E-16}\fluxunits}, corresponding to \mbox{SFR$_{\rm{Ly\alpha}}\approx\ $5.0$\pm$0.5 \SFRunits}. The velocity width is \mbox{$\Delta v$(\LAEc) $\approx 400$ \kms}, and equivalent width \LAEcEW, consistent with star formation being the primary driver of Ly$\alpha$ emission. We also note a coherent absorption feature at $\sim -400$\kms\ in spectra from at least three objects; the QSO, \LAEa\ and ``\LAEb" which could imply the presence of an expanding neutral gas shell with an extent of at least $24$ pkpc. \\
\end{abstract}


\section{Introduction} \label{sec:intro}

The \BR\ system is a prime example of a gas-rich major-merger at high redshift, which theory and simulations suggest are key to our understanding of galaxy evolution. Quasi-Stellar Object (QSO) `\BR', was discovered in the {\emph{APM-BRI}} survey \citep{IrwinMcMahonHazard1991} -- it was the first object at z$>4$ to be detected in CO emission \citep{Ohta96}, and these molecular gas observations revealed for the first time an optically-obscured sub-mm galaxy (SMG) lying $\sim 4 \,\arcsec$ to the North-West of the QSO (\citealt{Omont96}, \citealt{Iono2006}). Subsequently, the field has been targeted with a multi-wavelength campaign of observations to measure its gas content (\citealt{Wagg2012}, \citealt{Carilli2013}), and chemical abundances (\citealt{Decarli2018}, \citealt{Lehnert20}). 

Indeed, the hierarchical growth of structure predicts that QSOs at high redshift are biased tracers of galaxy formation, situated in overdense environments (e.g. \citealt{Overzier09}). In the case of QSO \BR, in addition to the nearby SMG, narrow-band data have indicated the presence of two Lyman-$\alpha$ emitters (LAEs) in the vicinity (\citealt{Hu1996}, \citealt{Omont96}, \citealt{Ohyama2004}, \citealt{Salome2012}): the source denoted \LAEa, positioned to the North-West of the QSO (in the direction of the SMG) and \LAEb\ towards the South-West. Futhermore, \CII$_{158\rm{\mu m}}$ observations of the entire \BR\ field from the commissioning of the Atacama Large Millimetre Array (\ALMA; \citealt{WootenThompson09}), suggest the possible presence of a bridge of \CII$_{158\rm{\mu m}}$ emission between the QSO and the SMG, tracing cooler ionised and/or neutral gas, intriguingly with an indication of a local maximum at the position of \LAEa. Together, this makes \BR\ an ideal system to study a diverse population of galaxies evolving in one of the most overdense regions of the Universe known, just $\sim 1.2$ Gyr after the Big Bang.

The nature of the two companion objects denoted \LAEa\ and \LAEb\ has been debated. The object that appears in \HST\ and \HSTb\ imaging, later named \LAEa, was spectroscopically confirmed by the detection of Ly$\alpha$ emission in \cite{Hu1996}. The optical line ratios presented in \cite{Williams14} suggest that the primary source of Ly$\alpha$ emission in \LAEa\ is star formation i.e. neither \CIV\ nor \HeII\ emission is detected, which should each be present in the case of photoionisation by an active galactic nucleus (AGN)\footnote{Although \CIV\ could be suppressed in a low metallicity system, \HeII\ should be detectable regardless, and thus places a strong constraint on the powering mechanism of the Ly$\alpha$ emission.}. \LAEa\ is detected in \CII$_{158\rm{\mu m}}$ (\citealt{Carilli2013}, \citealt{Wagg2012}), and shows a narrow emission line (of width $\Delta$v$_{\rm{[CII]}}$ = 56 \kms\ at full width half-maximum; FWHM), taking this as an indicator of systemic redshift, the Ly$\alpha$ emission is offset by 49 \kms\ from its predicted position in wavelength \citep{Williams14}. Observations in \cite{Decarli2018}, \cite{Pavesi16}, and \cite{Lee19} detect \NII$_{122\rm{\mu m}}$ emission from ionised Nitrogen at the position of \LAEa, at levels which suggest an origin within \HII\ regions.

\begin{figure*}
    \centering
	\includegraphics[width=0.95\columnwidth]{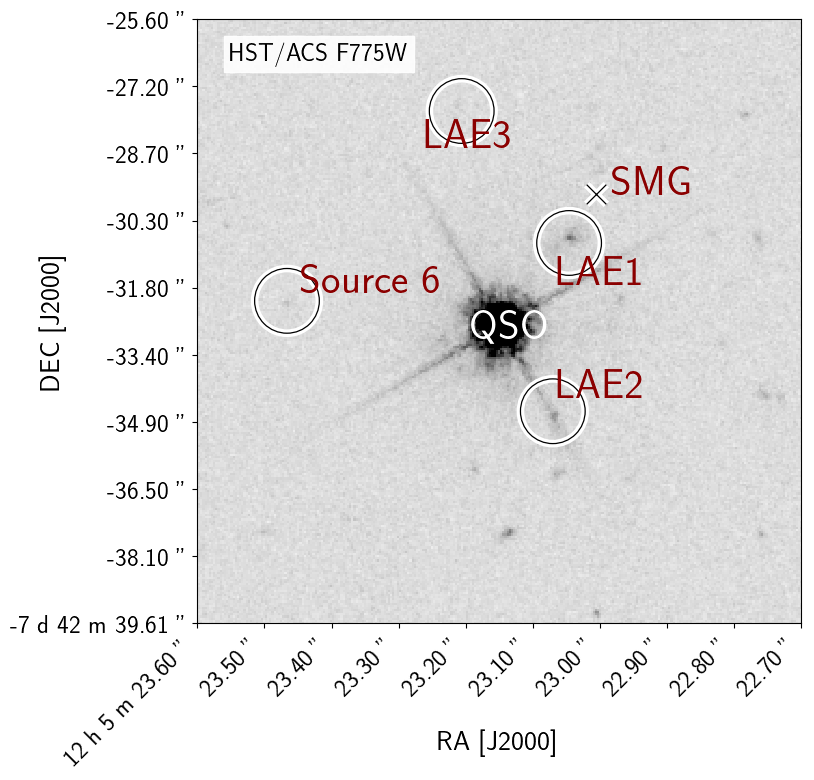}
	\includegraphics[width=0.95\columnwidth]{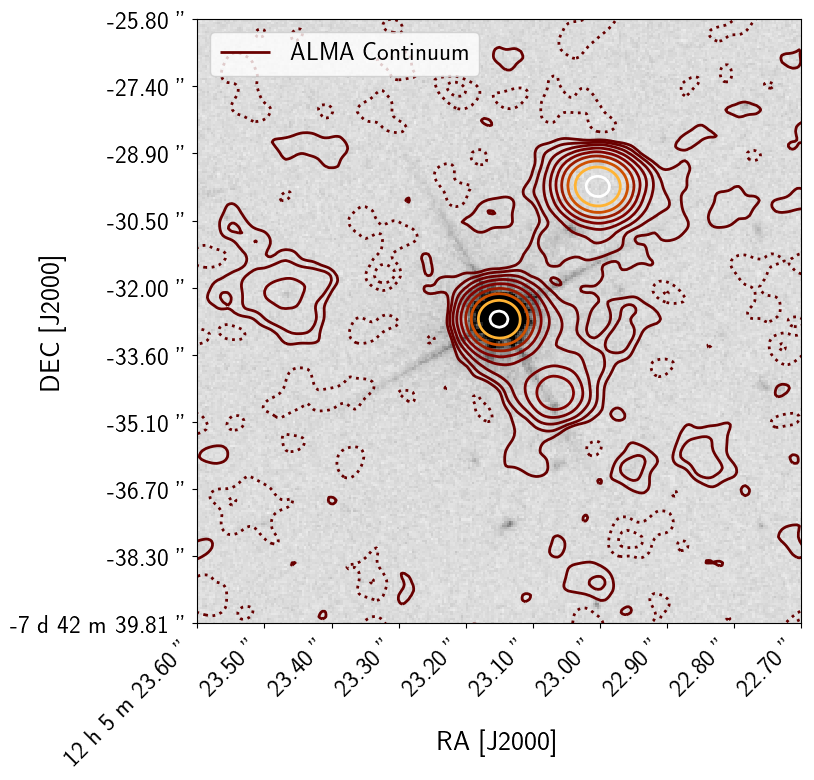}
	\caption{\HST\ imaging is displayed in both panels, providing the highest resolution optical data available for the field. In the left-hand panel we highlight the positions of the QSO, the (optically-obscured) SMG, two known objects dubbed \LAEa\ and \LAEb; a previously unreported source that we present in this work; \LAEc, and another previously unreported source which we name Source 6; hereafter "\LAEd". In the right-hand panel we display the same HST image, this time overlaid with sub-mm continuum contours from \ALMA. Contours are linearly spaced at $\pm$ $1.5, 3.0, 6.0, 12.0, 24.0, 48.0, 96.0, 192.0$ and $384.0\,\sigma$ where $1\sigma=12.38~\mu$Jy\,beam$^{-1}$, and negative contours are represented by dotted lines. The presence of the QSO and the optically-obscured SMG are both demonstrated at $>\!\!15\,\sigma$, with emission of lower-significance present surrounding the positions of \LAEa, \LAEb, and \LAEd.}
    \label{Fig: HST}
\end{figure*} 

\LAEb\ (\citealt{Hu97}, \citealt{Salome2012}), appears in narrowband imaging to be an LAE at the redshift of the QSO. Long-slit spectroscopic observations in \cite{Williams14} confirmed the presence of Ly$\alpha$ emission at the position of \LAEb, at approximately the redshift of the QSO, lending support to the idea that the object was indeed an LAE associasted with this group. The \CII\ emission from \LAEb\ falls at the edge of the \ALMA\ spectral setup, and hence it is not easy to judge the peak frequency or velocity width of the line (see \citealt{Carilli2013} and \citealt{Wagg2012}). \cite{Decarli2018} however reported an {[N{\sc{ii}}]}/\CII\ ratio for this object, which indicated that this sub-mm emission in \LAEb\ is likely to originate from \HII\ dominated regions, and as such the object could be forming stars. \\

In addition to the presence of companion galaxies at the redshift of the QSO, these massive objects are predicted to reside at the nodes of large-scale structure, composed of sheets and filaments of HI gas, known as the `cosmic web' (e.g. \citealt{springel06}). This gas is too diffuse to form stars, but instead is funnelled along the filamentary structure onto massive dark-matter halos hosting the QSO and/or other massive collapsed objects (e.g. \citealt{vandevoort11}), acting as fuel for their star formation. As a QSO is `fed' by this cool gas ($T \sim 10^4$ K), a number of physical processes (whose relative contributions are debated) lead to the emission of Ly$\alpha$ photons which, due to their resonant nature in HI gas, require careful interpretation, including modelling of their complex radiative transfer processes (e.g. \citealt{Michel-Dansac20}). Although extended Ly$\alpha$ halos can arise surrounding a diverse set of objects (e.g. \citealt{Venemans07}) in the case of a central QSO, the prime candidates responsible for powering the Ly$\alpha$ emission could be any of the following; (A) Photoionisation of the cool gas by a centrally located AGN (sometimes referred to as `Ly$\alpha$ fluorescence'; \citealt{Cantalupo10}, \citealt{Prescott15}); (B) `gravitational cooling' of in-flowing (pristine) gas, in which collisional excitation of atoms is the dominant power source (\citealt{Smith07}; \citealt{Rosdahl12}; \citealt{Daddi20}); and (C) shock-heating of the gas as a result of violent, possibly jet-induced star formation (e.g. \citealt{Taniguchi01}). In addition to these processes, star formation within the QSO's host galaxy, and/or star formation within other satellite galaxies could act as energy sources to also photoionise some fraction of the cool gas.\\

In recent years, observations from the panoramic integral field spectrograph \MUSE\ \citep{Bacon2010} have revolutionised the field of study surrounding the detection and analysis of extended Ly$\alpha$ halos/nebulae around QSOs and `normal' star-forming galaxies (\citealt{Bacon2015}, \citealt{Wisotzki16}, \citealt{Bacon2017}, \citealt{Inami2017}, \citealt{Drake2017a}, \citealt{Drake2017b},  \citealt{Leclercq2017}). The high spectral ($\Delta\lambda=1.25$ \AA) and spatial (0.202\,\arcsec) resolution have revealed detailed spatially-resolved kinematic maps of Ly$\alpha$ halos surrounding QSOs at the highest redshifts (\citealt{Farina2017}, \citealt{Ginolfi2018},  \citealt{Drake2019}, \citealt{Farina19}), and at z$\sim$2-3 where other rest-frame ultraviolet emission lines are accessible with \MUSE\ e.g. \CIV\ and/or \HeII\, potentially enabling constraints on the powering mechanisms of the halos (\citealt{ArrigoniBattaia15a}, \citealt{ArrigoniBattaia15b},
\citealt{Borisova2016}, \citealt{ArrigoniBattaia2018}, \citealt{Marino2019}; Also see results from the Keck Cosmic Web Imager e.g. \citealt{Cai19}).\\

Until now, observations of \BR's ionised gas content, traced by Ly$\alpha$ emission, have been limited to photometry from broad or narrow-bands, long-slit spectroscopy, and early IFU observations from TIGER \cite{Petitjean96}. In this paper we present IFU data covering the \BR\ field from \MUSE\ -- simultaneously revealing an extended Ly$\alpha$ halo around the QSO, allowing the re-analysis of Ly$\alpha$ emission from companion galaxies embedded within the halo, and comparing the ionised- and cool-gas properties of both Ly$\alpha$ halo and companion objects in the field. \\

This paper proceeds as follows; in Section \ref{sect:obs} we describe the data used in this paper, and its processing before analysis. The data consist of archival HST imaging, an archival \ALMA\ \CII\ datacube, deep \ALMA\ dust continuum imaging, and finally the \MUSE\ datacube. We also describe briefly here our method for PSF-subtraction in the \MUSE\ cube. In Section \ref{sect:res-Lya-halo} we present \MUSE\ images of the field, and a spectrum of the QSO, followed by the results of our PSF-subtraction. Here we analyse the spatial extent and morphology of the Ly$\alpha$ halo, and take advantage of the spatial resolution of \MUSE\ to produce moment maps of the Ly$\alpha$ emission, and search for overlap or coincidence of Ly$\alpha$ and \CII\ emission across the field. We speculate on the dominant powering mechanism of the Ly$\alpha$ halo and perform a search for extended \CIV\ emission. Next, in Section \ref{sect:res-LAEs} we present images and spectra of a series of companions in the field, including the SMG, \LAEa\ and \LAEb. For each object in the system we assess the Ly$\alpha$ emission and derive velocity widths, star-formation rates (SFRs; where appropriate) and constraints on the rest-frame equivalent widths of Ly$\alpha$ (EW$_0$; again where possible) before using these measurements to re-assess the nature of the proposed companions. We summarise our findings on the QSO's Ly$\alpha$ halo and all accompanying objects in Section \ref{sect:concl}.\\

We assume a $\Lambda$CDM cosmology with $\Omega_m = 0.3$, $\Omega_{\Lambda} = 0.7$ and H$_0=70$ km s$^{-1}$ Mpc$^{-1}$. In this cosmology, \mbox{1\,\arcsec\ = 6.64 pkpc} at z $\approx$ 4.69.\\

\begin{figure*}
    \centering
	\includegraphics[width=0.99\textwidth]{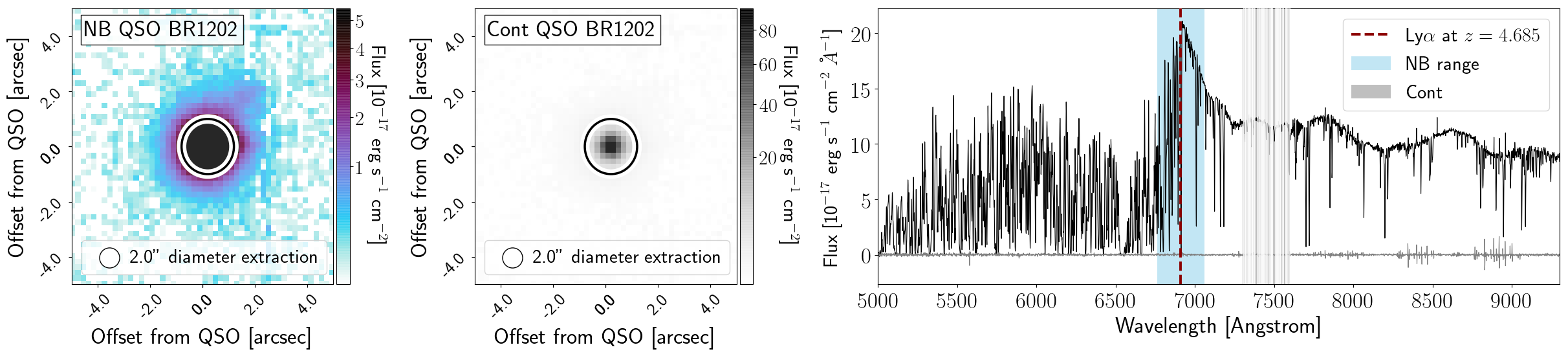}
	\caption{MUSE images and spectra of QSO \BR. The left-hand panel shows a fixed-width narrow-band image of $\Delta \lambda = 120$ \AA\ centred on the predicted wavelength of Ly$\alpha$ according to the QSO's \CII\ redshift from \ALMA. The wavelength range of the narrow-band image is highlighted in orange on the spectrum to the right. In the central panel we show the PSF image (see text for details) with the spectral layers used to construct the image highlighted in pink on the spectrum. In the final column, we show the QSO spectrum extracted in an aperture of 2 arcseconds in diameter, the extraction region is given by the black circle on the narrow-band and PSF images. A dashed red line on the spectrum gives the predicted peak wavelength of the Ly$\alpha$ emission according to the systemic redshift of the QSO.}
    \label{Fig: PSF}
\end{figure*}

\section{Observations and Data Reduction} \label{sect:obs}

In addition to our analysis of the new \MUSE\ observations, we make use of two existing datasets from \ALMA; allowing a multiwavelength comparison of the ionised and neutral gas content of the \BR\ system; and archival HST imaging to give an optical overview of the field at the highest-resolution available. The datasets are described below.

\subsection{HST overview of the field} \label{sect:obs-HST}

In Figure \ref{Fig: HST} we show the archival \HST\ image of \BR\ highlighting in the left panel the positions of the QSO, the SMG, two known-LAEs (\LAEa, \LAEb) and an additional LAE discovered in this work; \LAEc. In the right-hand panel we show the \HST\ image again, overlaying sub-mm dust continuum contours from the deepest ALMA observations of the field, see Section \ref{sect:obs-ALMA cont} for a description. Both the archival HST imaging, and the dust-continuum maps are used throughout this work for orientation purposes only. \\

\subsection{ALMA Observations} \label{sect:obs-ALMA}

\subsubsection{Deep Continuum Imaging}\label{sect:obs-ALMA cont}
In order to obtain a deep mm continuum map we explored all public data in the ALMA archive available for our system. Projected baseline lengths for the combined continuum dataset are between 14 - 1440 m, with the 80th percentile at 288 m. We decided to combine the data in the frequency range of 190 -- 297~GHz (bands 5, 6 and 7). Inside this 107~GHz bandwidth, 33~GHz was observed with six frequency setups\footnote{Few band 6 setups partially overlap.}. These setups cover several far-infrared (FIR) bright molecular emission lines of CO (high excitation rotational CO lines J$_{up}$ = 10, 12, 14) and H$_2$O, and a fine atomic transition line \NII\ at $205$ microns. A total width of 1500\,km s$^{-1}$ centered at each of the emission lines was excluded from the continuum imaging process. This width choice corresponds to two times the FWHM of the brightest \CII\ line detected in the SMG inside this system \cite{Carilli2013}. The continuum map was obtained from the remaining line-free channels spanning an effective bandwidth of 25 GHz. Individual spectral setups were observed at roughly similar resolutions, with synthesized beam FWHM range of $0.6$\,\arcsec - $1.5$\,\arcsec, and a position angle of 88$^{\circ}$, thus allowing joint imaging of all the datasets. For imaging we use the {\sc tclean} task contained in the Common Astronomy Software Applications (CASA\footnote{CASA version 5.4.0-70}) package \citep{mcmullin07}. Given the large bandwidth available, we image the data using the multi-term multi-frequency synthesis (MTMFS, \textsc{nterms=2}, \citealt{rau11}). The data were imaged with natural weighting to maximize the point source sensitivity. The synthesised beam side-lobes of the combined dataset are smaller than 5\%. Due to good quality of the final map, no additional weighting of visibilities was deemed necessary\footnote{We have checked that re-weighing of combined visibilities using the STATWT task in CASA does not further improve the map quality.}. Cleaning was performed with the multi-scales algorithm (using scales corresponding to a single pixel, 1x, and 3x the beam size)  first down to $5\sigma$ in the entire map, and then further down to $1.5\sigma$ inside manually defined cleaning regions, which outline the observed emission. The final continuum map is given at the monochromatic frequency of 243.5~GHz, resolution of $0.83\,\arcsec \times 0.73\,\arcsec$, and has a root-mean-square (rms) noise level of $1\sigma=12.38~\mu$Jy\,beam$^{-1}$.

\subsubsection{Archival [CII] Observations} \label{sect:obs-ALMA cii}
We make use of \ALMA\ $335$ GHz (Band $7$) Science Verification data with a central frequency targeting the \CII\ line at the redshift of the QSO and the SMG. The data were first presented in \cite{Wagg2012} and \cite{Carilli2013}. We applied a velocity-frame correction to the published data to convert from the observed frame (topocentric) to the local standard of rest (LSRK) for accurate comparison to velocities in the \MUSE\ datacube. 

\subsection{MUSE Observations} \label{sect:obs-MUSE}

\subsubsection{MUSE Data Reduction} \label{sect:obs-MUSE DR}

\MUSE\ data were taken as part of ESO programme $0102.A-0428(A)$, PI Farina, and reduced as in \cite{Farina2017} and \cite{Farina19} using the \MUSE\ Data Reduction Software version 2.6 (\citealt{Weilbacher12}, \citealt{Weilbacher14}). Two exposures of 1426 s were taken, with $<$ $5\,\arcsec$ shifts and 90 degree rotations. The PSF has a size of 0.6$\,\arcsec$ at the observed wavelengths of the Ly$\alpha$ line (the median delivered PSF on stars in the field is closer to 0.7$\,\arcsec$). The 5$\sigma$ surface brightness detection limit is $4.2$ $\times$ $10^{-17}$ \SBunits\ for an aperture of 1 square arcsecond. Data have been corrected for galactic extinction, and emission from night sky lines is removed using the Zurich Atmospheric Purge software (ZAP; \citealt{Soto2016}).

\begin{figure*}
    \centering
	\includegraphics[width=0.95\textwidth]{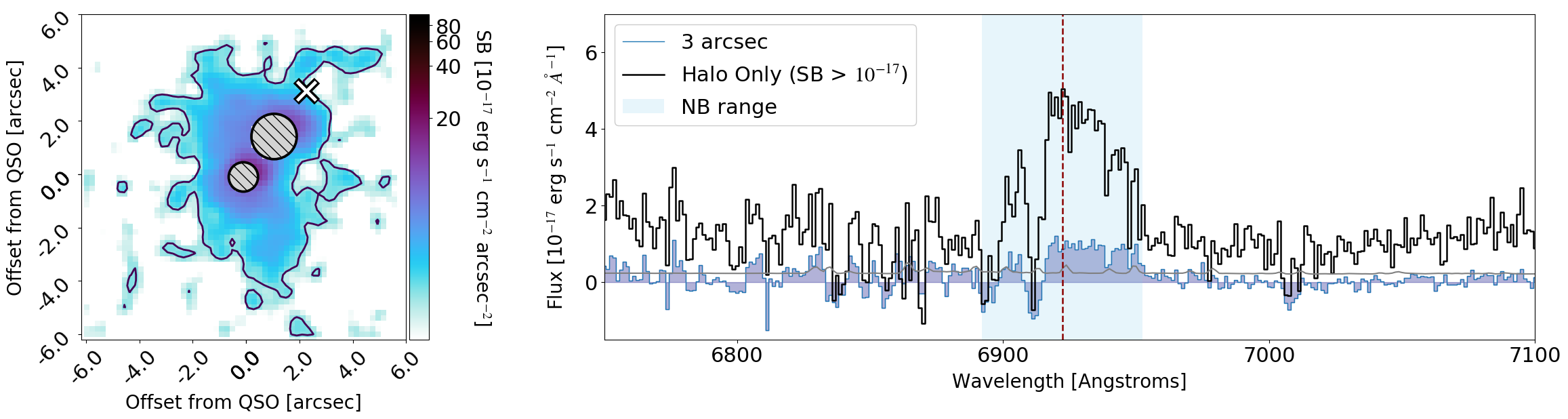}
	\caption{A smoothed surface brightness image and spectrum of the continuum- and PSF-subtracted Ly$\alpha$ halo surrounding the QSO. In the left-hand panel we show a narrow-band image smoothed with a Gaussian kernel of $\sigma = 1$ pixel, in units of surface brightness. A single contour is displayed at  SB$=$ $\num{1e-17}$ \SBunits\ (black line). The QSO position is masked over a diameter of 1$\,\arcsec$, and \LAEa\ is masked over a diameter of 1.6$\,\arcsec$, we also mark the position of the SMG on this image with a white cross. In the right-hand panel we show in filled-blue the spectrum extracted from the continuum- and PSF-subtracted cube within a 3$\,\arcsec$ diameter aperture (after employing the mask for residuals), and overlay in black a spectrum extracted by summing all the voxels (volumetric pixels) lying within the $\num{1e-17}$ \SBunits\ contour with the QSO and the position of \LAEa\ masked. The dashed red line gives the predicted position of the peak of the Ly$\alpha$ emission according to the systemic redshift of the QSO (Table \ref{tab:objs}), the shaded orange region corresponds to the wavelength region that makes up the narrow-band image, and we also overplot in red the sky spectrum.}
    \label{Fig: Blob Spec}
\end{figure*}

\begin{figure*}
    \centering
	\includegraphics[width=0.49\textwidth]{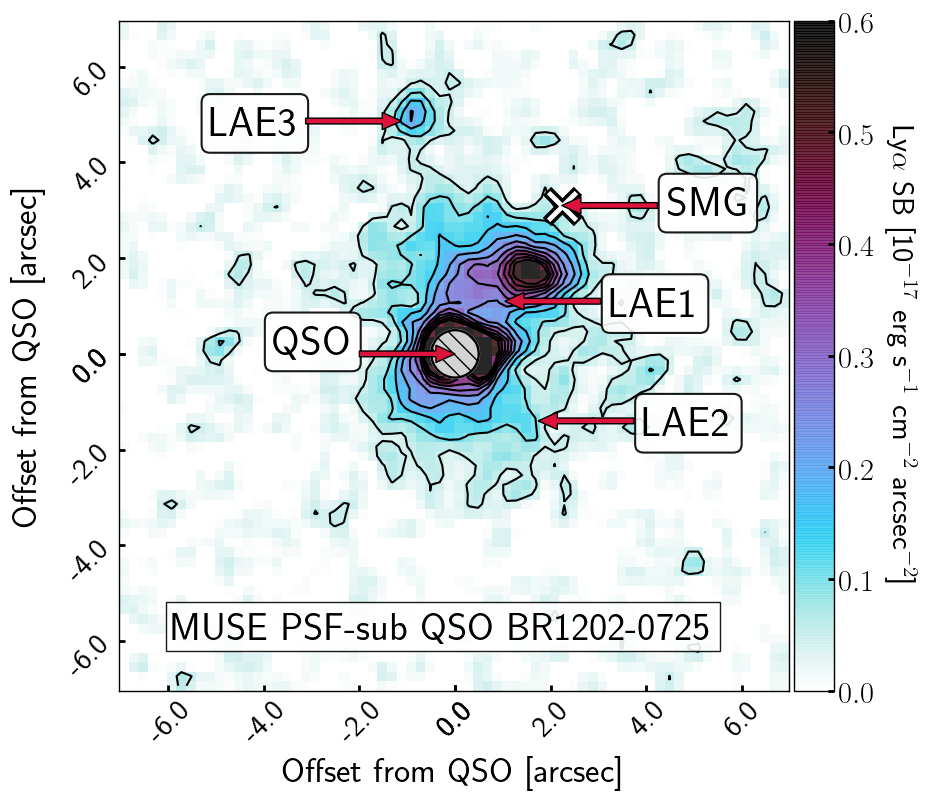}
	\includegraphics[width=0.49\textwidth]{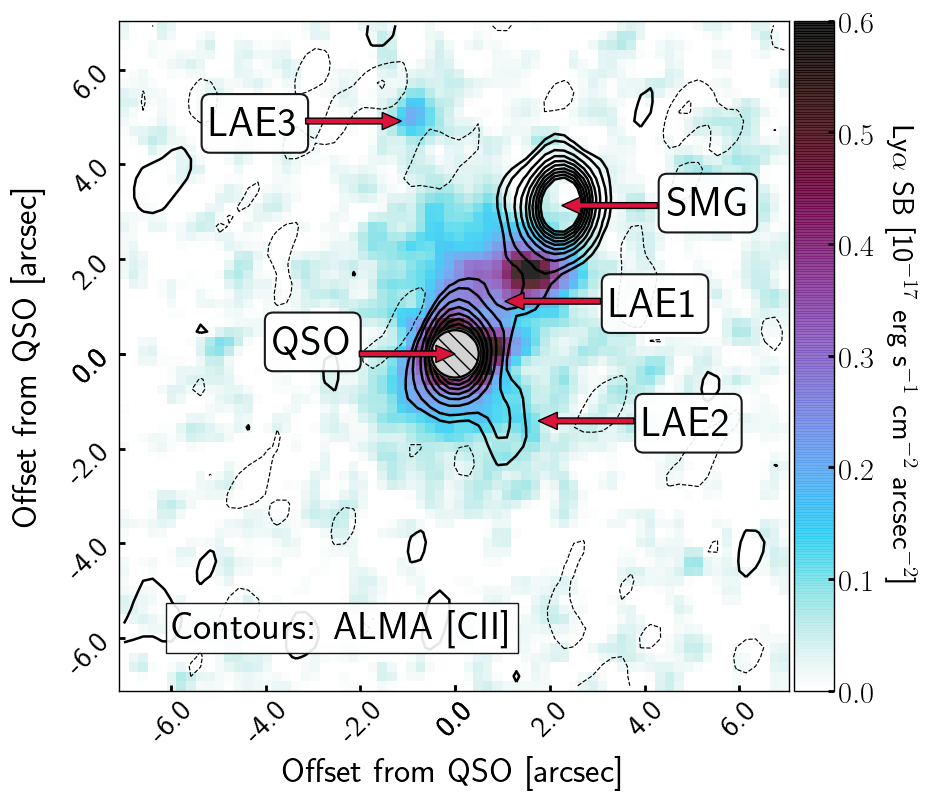}
	\includegraphics[width=0.95\textwidth]{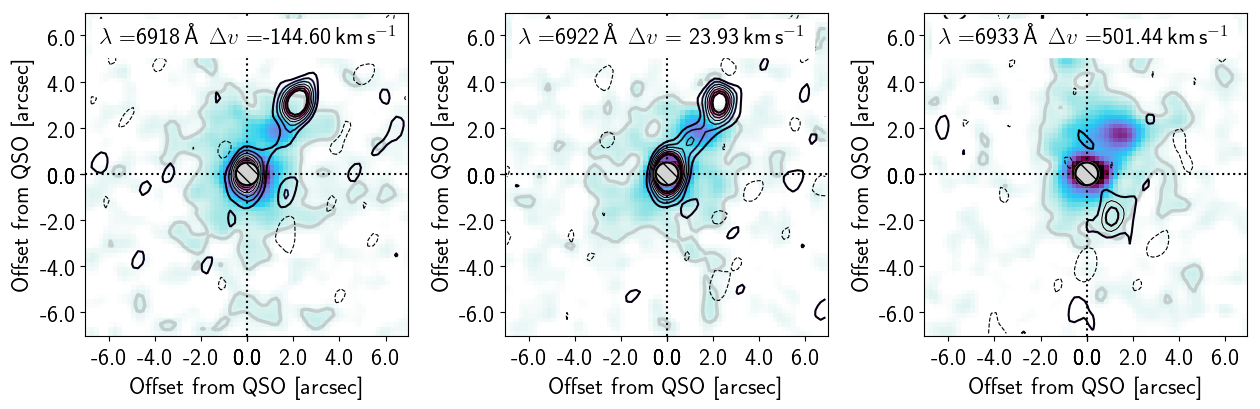}
	\caption{PSF-subtracted narrow-band \MUSE\ images centred on the Ly$\alpha$ line are displayed together with contours to depict either the significance of the Ly$\alpha$ emission or the coincident \CII\ emission. In the upper panels the \MUSE\ image is a collapse of the monochromatic  wavelength  layers  between  $6927$ and $6932$ \AA; i.e. a narrow velocity range ($\sim200$ km\,s$^{-1}$) in surface brightness units (\SBunits). In the left-hand panel we overlay contours ranging between $\num{5e-19}$ and $\num{5e-18}$ \SBunits\ denoting the Ly$\alpha$ surface brightness. In the right-hand panel we instead overlay linearly-spaced contours denoting \CII\ emission from archival \ALMA\ observations across the entire velocity range of the data cube. Contours are linearly spaced at $\pm$ $1.5, 3.0, 6.0, 9.0, 12.0$ and $15.0\,\sigma$, and negative contours are represented by dotted lines. In both panels residuals from PSF subtraction are masked across $1\,\arcsec$ at the position of the QSO. In the lower three panels we show cutouts at specific velocities of Ly$\alpha$ and \CII\ emission taken from the full set of channel maps presented in Appendix B. In these panels we include a single grey contour to depict the 0.5 $\sigma$ level of the extended Ly$\alpha$, and overlay linearly-spaced contours denoting \CII\ emission beginning at 2$\sigma$.}
    \label{Fig: NB}
\end{figure*}

% Obj table
\begin{table*}
	\centering
	\caption{Known objects in the BR1202 system at the time of writing. The first column gives the literature redshift oftheobject. The second column gives the offset in velocity between this object and the QSO. The third column gives the predicted wavelength of Ly$\alpha$ emission given the systemic redshift of the object. In the final column we include notes on each object. }
	\label{tab:objs}
	\begin{tabular}{lcccr} % columns, alignment for each
		\hline \hline
{\bf{Object}} &	{\bf{$z$$_{{\rm{[CII]}}}$}}\footnote{\cite{Carilli2013}, formal errors from Gaussian fitting on the redshifts are each \textless 0.0003.} &	{\bf{Vel Offset}}\footnote{Object's velocity offset from the QSO, where negative numbers represent a blue-shift, and positive numbers a red-shift.} & {\bf{Pred $\lambda$ Ly$\alpha$}}  & {\bf{References}}     \\

 	& & \multicolumn{1}{C}{\rm{[kms$^{-1}$]}}  &  [\AA] &\\
\hline
{\bf{QSO}} &4.6942 & 0.0 & 6922.27 & \cite{McMahon94}, \cite{Isaak1994} \\ 
{\bf{SMG}} & 4.6915 & -142.1 &6918.99 & \cite{Omont1996}, \cite{Riechers06} \\ 
{\bf{\LAEa}} & 4.6950 & 42.0 & 6923.24 & \cite{Hu1996}, \cite{Ohyama2004} \\
{\bf{\LAEb}} & 4.7055 & 595.5 & 6936.01  & \cite{Hu97}, \cite{Wagg2012}, \cite{Carilli2013} \\

	\hline
	\end{tabular}
\end{table*}

\subsubsection{MUSE PSF Subtraction} \label{sect:obs-MUSE psf}

With a view to uncovering low-surface-brightness Ly$\alpha$ emission in the MUSE data surrounding QSO \BR, we follow the same procedure as in \cite{Drake2019} to model and subtract the point-spread function (PSF) in the data (also demonstrated in \citealt{Farina19}). In brief, this entails collapsing several spectral layers of the QSO's 
optical continuum which is dominated by light from the accretion disk of the AGN, and appears as a point source at the resolution of MUSE. By using the quasar itself to create the ``PSF image" we avoid issues of spatial PSF-variation/interpolation across the field. The wavelength layers chosen to construct the PSF image are highlighted in pink in the right-hand panel of Figure \ref{Fig: PSF}. We then work systematically through the MUSE cube, scaling our PSF image such that the flux in the pixel at the image peak becomes equal to the flux of the QSO in the same spatial pixel. By subtracting this scaled PSF image from each wavelength layer we produce an entire PSF-subtracted datacube. Finally, as in \cite{Drake2019}, we mask an ellipse on every layer of the datacube of radii equal to the FWHM of a 2-dimensional Gaussian fit to the PSF image, and exclude this region from further analysis to avoid residuals near the bright central source, unless otherwise noted.

\section{Extended Ly$\alpha$ in the \BR\ field} \label{sect:res-Lya-halo}

\subsection{Total Flux of Ly$\alpha$ Halo around QSO \BR} \label{sect:totflux}
In Figure \ref{Fig: Blob Spec} we show a narrow-band image comprised of all the Ly$\alpha$ emission in a square 12\,\arcsec\ region surrounding the QSO. The narrowband width was chosen to encompass the entirety of the red side of the Ly$\alpha$ line after PSF-subtraction, and to include emission out to the same velocity bluewards of the systemic redshift of the QSO. This amounts to a total of 61\AA, or $\approx 2640$ \kms. In the right-hand panel of Figure \ref{Fig: Blob Spec} we show the PSF-subtracted spectrum extracted  in a 3\,\arcsec\ diameter aperture shaded in blue, and over plot with a black line the PSF-subtracted spectrum that results from summing all emission within the $\num{1e-17}$ \SBunits\ surface-brightness contour. Interestingly the halo's spectral shape is somewhat `flat-topped', which may be the result of contributions from objects  at different velocities within the halo, or simply represent an intrinsically broad line. The flux of diffuse Ly$\alpha$ within the {\mbox{SB\,$=\,\num{1e-17}$ \SBunits}}\ surface-brightness contour is \haloflux\, \fluxunits, after continuum-subtraction, masking both the position of the QSO (across a $1\,\arcsec$ diameter), and \LAEa\ (across a $1.6\,\arcsec$ diameter). If the halo emission were powered solely by star formation, we could translate this to an SFR according to Equation \ref{eq:1} \citep{ouchi03}:

\begin{equation}
\label{eq:1}
    \mathrm{SFR}_{\mathrm{Ly\alpha}} (\mathrm{M}_{\odot} yr^{-1}) =  \mathrm{L}_{{\rm{Ly\alpha}}} \times 1.05 \times 10^{42.0} \mathrm{erg\, s}^{-1}
\end{equation}

\noindent where L$_{{\rm{Ly\alpha}}}$ is the Ly$\alpha$ luminosity in cgs units. Our measured flux translates to a luminosity of {\mbox{L$_{\rm{Ly\alpha}}\,=$ $\num{3.11E+44}$ \lumunits}}, which would then correspond to an SFR of almost $\approx 300$ \SFRunits.\\

\subsection{Morphology of Extended Ly$\alpha$ around QSO \BR} \label{sect:morph}
 To examine the distribution of the diffuse Ly$\alpha$ in more detail, we show in Figure \ref{Fig: NB} a narrow velocity range ($\sim\,200$ \kms; i.e. a collapse of the monochromatic wavelength layers between $6927$ and $6932$ \AA). This choice of velocity range reveals filamentary structure at low surface-brightness levels, highlighting the complex morphology of the halo, while encompassing wavelength layers within which the QSO, \LAEa, \LAEb\ and \LAEc\ are all visible. In the top left-hand panel we show the Ly$\alpha$ surface brightness, contoured between SB\,$=\,\num{5e-19}$ and {\mbox{$\num{5e-18}$ \SBunits}}, and again highlight the positions of the QSO, the SMG, and three LAEs in the system. Figure \ref{Fig: NB} demonstrates that diffuse Ly$\alpha$ connects all three objects, while none is seen surrounding the SMG. Interestingly, the low-surface-brightness emission in Ly$\alpha$ extends directly in the direction of the optically-obscured SMG, and fully encompasses the position of \LAEa, however by the position of the SMG, the halo is no longer visible in this velocity range. The Ly$\alpha$ emission at the position of \LAEb\ however appears indistinguishable from the halo extending from the QSO in this image -- we return to this in Section \ref{sect: nature of objs}. Finally, \LAEc\ appears as a distinct source in Ly$\alpha$, however its emission is possibly connected via a low-surface-brightness bridge of Ly$\alpha$ emission to the position of the QSO. In the top right-hand panel we display the same Ly$\alpha$ surface-brightness image, but this time overlay contours depicting \CII\ emission (\citealt{Wagg2012} and \citealt{Carilli2013}) from the entire collapsed data cube. In the lower panels, we show cutouts at three different velocites relative to the systemic redshift of QSO \BR\ (i.e. a velocity slice of the Ly$\alpha$ halo overlaid with contours depicting \CII\ emission in the corresponding velocity slice). The velocities shown here were selected from the full series of channel maps presented in Appendix B, and are chosen to highlight a number of features. First, the purported \CII\ bridge between the QSO and the SMG. This is seen at negative velocities (which correspond to the slightly lower redshift of the SMG than the QSO) which is seen across multiple channels. Secondly, in the central panel we show the channel closest to systemic, the extended \CII\ persists, with a local maximum thought to define the position of \LAEa's ISM. In the final panel, $500$ \kms\ red-ward of the QSO, the position of \LAEb\ becomes clear. Together, these data begin to highlight the diversity of properties of the objects in this field. While the QSO appears bright in both Ly$\alpha$ and \CII\ emission, the SMG appears only at sub-mm wavelengths. LAEs 1 and 2 show \CII\ emission most visible in channels at their respective velocities, however \LAEc\ does not show any associated \CII\ emission. We will return to each of these features and discuss the objects' nature in Section \ref{sect:res-LAEs}.

\begin{figure*}
    \centering
	\includegraphics[width=\textwidth]{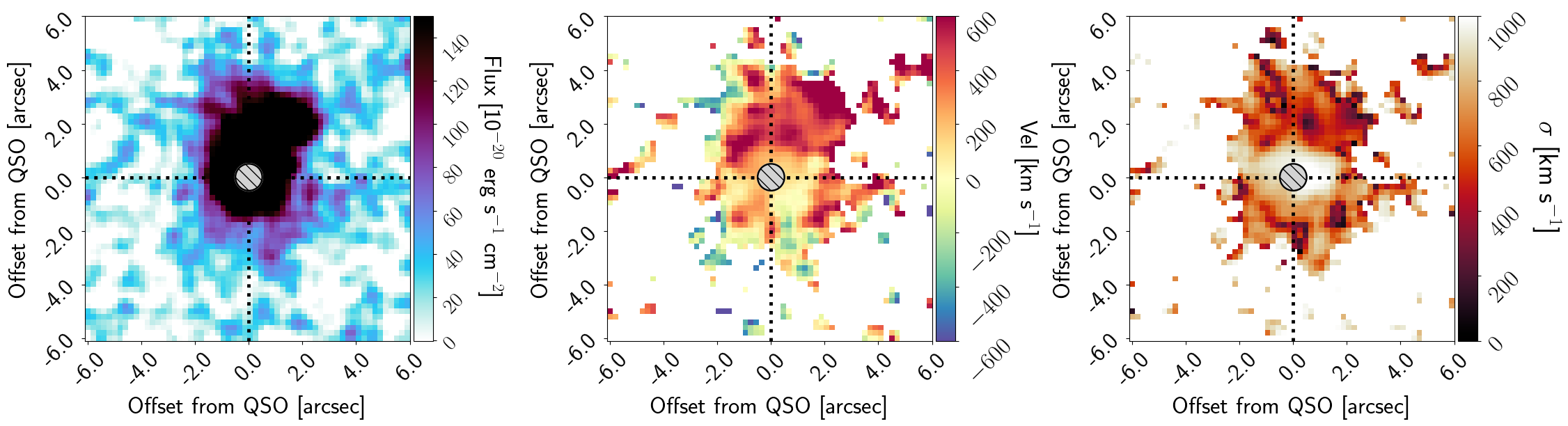}

	\caption{Ly$\alpha$ moment maps for the PSF-subtracted Ly$\alpha$ halo around QSO \BR. The left-hand panel shows moment zero (a flux image) containing all volumetric pixels, that make up the rest of the kinematic analysis. The moment analysis is performed across the wavelength range $6882-6962$ \AA - i.e. approximately the same range shown in Figure \ref{Fig: Blob Spec}. In the central panel we show the first moment, i.e. the velocity of each spatial pixel relative to the peak of the halo Ly$\alpha$ emission, and in the right-hand panel we show the second moment, or velocity dispersion in each spatial pixel. Residuals from the PSF subtraction are masked across $1\,\arcsec$ around the QSO in each panel.}
    \label{Fig: Lya moms}
\end{figure*}

\subsection{Kinematic analysis of Ly$\alpha$ around QSO \BR} \label{sect:kinLya}

To analyse the internal kinematics of the extended Ly$\alpha$ emission, we present zeroth-, first- and second-moment maps (representing the total flux, velocity and dispersion maps, respectively) in Figure \ref{Fig: Lya moms}, relative to the predicted peak of the Ly$\alpha$ line at a systemic redshift of $z = 4.6942$. We analyse the data in a manner consistent with \cite{Drake2019}; we first smooth the datacube in the two spatial directions with a Gaussian kernel of $\sigma = 1.0$ pixel, and calculate the non-parametric moments of the data, i.e. we do not assume any functional form for the Ly$\alpha$ spectral shape. In the first panel we show the flux-weighted zeroth-moment. This image/map is essentially the same as that shown on the left-hand side of Figure \ref{Fig: Blob Spec}.\\

In the central panel we show the first moment of the halo's Ly$\alpha$ emission, which gives the flux-weighted velocity of the halo gas relative to the peak of the emission, applying a uniform post-processing S/N cut of 1.0 on the moment zero image. The velocity structure is clumpy and complex. The majority of the extended emission, which appears to the North of the QSO, is red-shifted by a few hundred  km s$^{-1}$, only small patches of emission ($\leq2 \,\arcsec$ on-sky) appear blue-shifted, and they each appear towards the South of QSO. \LAEa\ does not appear kinematically distinct from the halo. No obvious signs of a flow of gas between the two objects is seen in this map, however for a more thorough examination of the relative velocities of Ly$\alpha$ and \CII\ emission across the field of view we refer the reader to Appendix \ref{App:channel maps} where we show a series of `channel maps', displaying images of the extended Ly$\alpha$ emission seen with \MUSE, overlaid with \CII\ contours from \ALMA\ (see \citealt{Carilli2013} and \citealt{Wagg2012}). The spectral resolution of \MUSE\ at $\sim 7000$\AA\ corresponds to $\sim\Delta v = 100 $ kms$^{-1}$. The channel spacing in velocity of the \CII\ emission seen with \ALMA, is $\sim\Delta v = 35$ km s$^{-1}$. The maps show that the extended Ly$\alpha$ emission is to some extent co-spatial with the \CII, and peaks at the same velocities as the \CII\ bridge between \BR\  and the SMG proposed in \cite{Carilli2013}. The Ly$\alpha$ emission is however much broader than the \CII, and as such Ly$\alpha$ still appears bright in channels beyond the end of the \ALMA\ \CII\ coverage. \\

Finally, in the third panel of Figure \ref{Fig: Lya moms}, we show the second moment, $\sigma$, giving the velocity dispersion of the gas, and applying the same uniform S/N cut of 1.0 on the moment zero image. Velocities close-in to the QSO are very high, of order $\sim1000$ km s$^{-1}$, with a region of lower values (of order $500$ km s$^{-1}$) in the direction of \LAEa.\\

\subsection{Constraints on the dynamical mass of the halo}
\label{sect:powering}
Using the moment maps, we can roughly estimate the dynamical mass of this gas given a number of assumptions. For instance, in the moment zero flux image (Figure \ref{Fig: Lya moms}, left) where the entire width of the Ly$\alpha$ line is collapsed, emission stretches $\approx 8\,\arcsec$ North$-$South, and $\approx 8\,\arcsec$ East$-$West surrounding the QSO. We therefore take an average radius of $4\,\arcsec$, which corresponds to $\approx27$ pkpc at $z=4.69$. In the moment 1 map (Figure \ref{Fig: Lya moms}, centre) we see a gradient of approximately $v\approx\pm 300$ \kms\ across the halo. Then, if one assumes rotation of the gas is responsible for the velocity gradient, we solve for dynamical mass $M_{\rm{\,dyn}}$:\\

\begin{equation}
    M_{\mathrm{dyn}} = rv^2/G
\end{equation}

where $G$ is the gravitational constant, $r$ is the halo radius, and $v$ is the velocity range across this radius. As we have no information on inclination angle, we place a lower limit on the dynamical mass of the halo; {\mbox{$M_{\rm{\,dyn}} \geq$ \num{6E11} M$_{\odot}$}}.  
Even without any correction for inclination angle, this is significantly larger than the combined molecular gas mass of the QSO-SMG system reported in the literature ($\sim10^{11}$ M$_{\odot}$; \citealt{Omont1996}, \citealt{Riechers06}), and 2 orders of magnitude larger than the QSO's black hole mass, {\mbox{M$_{{\rm{BH}}}= \num{1.9e9}$ M$_{\odot}$}} (\citealt{Carniani2013}).

\subsection{Speculation on Ly$\alpha$ powering mechanism}

The powering mechanisms of Ly$\alpha$ halos at high redshift have long been debated. \BR\ is an example of a system where any one of the proposed mechanisms outlined in Section \ref{sec:intro} could naively be assumed the primary driver of the halo emission, or perhaps more likely, a complex mixture of processes are responsible. The QSO has an obscured star-formation rate in excess of 1000 \SFRunits\ \citep{Carilli2013}, and as such, in-situ star-formation could be responsible for the extended Ly$\alpha$ emission. Likewise, copious amounts of pristine gas are required to fuel this major merger which could give rise to gravitational cooling of the gas as it is funnelled onto the QSO. It is perhaps of some significance then that QSO \BR\ is accompanied by the SMG of similar gas mass, inferred SFR, and dust content \citep{Carilli2013}, but that displays no prominant Ly$\alpha$ halo. In the absence of any further diagnostics, this lends support to the idea that the Ly$\alpha$ halo is directly linked to the actively accreting SMBH in the QSO, and not to star formation.

Several studies in the literature have recently taken steps towards identifying the powering mechanisms of Ly$\alpha$ halos around QSOs through the use of additional diagnostic lines \citep{ArrigoniBattaia18}. Motivated by these studies, we take advantage of the spectral coverage of \MUSE\ to search for any extended \CIV\ emission surrounding \BR. The metal line \CIV\ indicates whether the Ly$\alpha$ emitting gas has been enriched (i.e. it orignated within the host galaxy) or is pristine (falling onto the halo for the first time). In Appendix \ref{APP: CIV} we show the \MUSE\ spectrum of QSO \BR\ again, and overlay a composite quasar spectrum \citep{Selsing16}, `red-shifted' to $z=4.6942$. We highlight on this spectrum the predicted observed wavelengths of various emission lines, in particular \CIV\ $\lambda_{em}=1549$\AA. We extract an image and spectrum exactly as for Ly$\alpha$ in Figure \ref{Fig: Blob Spec}, but this time centred on the predicted wavlength of \CIV. We detect no extended \CIV\ emission down to a surface brightness limit of \num{0.19E-16} \SBunits\ in a square arcsecond. Given however that metal lines in high-redshift QSOs are frequently observed to be blueshifted with respect to the systemic velocity, we repeat this exercise in the 8500-8700 \AA\ region where we see a subtle, broad `bump' in the QSO spectrum, but find no evidence for extended emission in this region either. Unfortunately these results do not place any strong constraint on the powering mechanism of this Ly$\alpha$ halo; i.e. in the absence of strong radio emission \CIV\ luminosities are so much fainter than Ly$\alpha$ emission (e.g. typically \CIV/Ly$\alpha$ $\leq$ $0.13$ in Ly$\alpha$ `blobs' where no definitive power source has been determined; \citealt{ArrigoniBattaia18}). Although radio continuum has been observed in both the QSO and SMG here, the emission is consistent with a weak AGN or extreme star formation (\citealt{YunCarilli02}, \citealt{Carilli2013}) and not the powerful high-redshift radio galaxies (HzRGs) that are known to exhibit significant extended \CIV\ ( e.g. \citealt{Matsuoka09}).

\section{Companions in the Field} \label{sect:res-LAEs}

 In this Section, we present new spectra from \MUSE\ of the two known LAEs in addition to \LAEc\ discovered in this work. We present measurements of the Ly$\alpha$ flux, velocity width, and rest-frame equivalent width (EW$_0$), and discuss the results below. In addition we investigate the existence of an additional source, dubbed  ``\LAEd", motivated by an alignment of dust continuum emission and a compact object seen in \HST\ imaging. The diversity of object properties (and their staggered discovery) means that the dataset in which each object's position is defined varies. Where possible, we take the object's position directly from table 1 of \citealt{Carilli2013} (i.e. for the QSO, SMG, \LAEa, and \LAEb). For the QSO, SMG and \LAEb\ this is the dust continuum position. For \LAEa\ this is the \CII\ position. \LAEc\ is defined by its \MUSE\ detection, and the position of \LAEd\ is defined as the centre of the dust emission.

\subsection{Ly$\alpha$ properties of companions}
\label{sect: Lya props}

In Table \ref{tab:Lya props} we summarise our measurements of Ly$\alpha$ emission from the \MUSE\ cube for each of the objects known in the \BR\ field, plus \LAEc, and the potential source `\LAEd'. We also include cutout images and spectra in Appendix \ref{App: flux measurements} to demonstrate our choice of aperture size, and the \MUSE\ spectrum from which we measure the Ly$\alpha$ flux (Table \ref{tab:Lya props} column 5).\\

The complexity of the \BR\ system and the diffuse Ly$\alpha$ emission in the field make it difficult to identify individual objects' Ly$\alpha$ in an image, as objects are essentially `embedded' within the diffuse Ly$\alpha$ halo stretching across the entire field. For this reason we choose very narrow velocity ranges over which we display the corresponding Ly$\alpha$ image in the final column of Figure \ref{Fig: Fluxes}. We then choose an aperture on these images to encapsulate the emission from a particular object.  \\

For two objects, \LAEa\ and \LAEc, the Ly$\alpha$ flux can be estimated simply by fitting a Gaussian profile to the 1D spectral extraction. For these two objects we also place constraints on the rest-frame equivalent width (EW$_0$) of Ly$\alpha$ from the MUSE spectrum. Accurate measurements of LAEs' EW$_0$ at this redshift are potentially of great interest due to their ability to distinguish between powering mechanisms of Ly$\alpha$ emission. Models based on standard IMFs, stellar populations and metallicity ranges for instance predict a maximum value of EW$_0 = 240$\AA\ for emission powered by star formation (see \citealt{Schaerer2003} and \citealt{Hashimoto17}). For \LAEb\ it is difficult to produce a good fit from the 1D spectrum, and so we first fix the peak of the Gaussian to the predicted wavelength of Ly$\alpha$ emission corresponding to \LAEb's systemic redshift. For the remaining objects (the QSO, the SMG, and `\LAEd') we simply sum the flux across a $50$ \AA\ window (i.e. $\approx$ twice the measured FWHM of \LAEa).\\

 We recover a Ly$\alpha$ flux estimate for \LAEa\ of {\mbox{F$_{{\rm{Ly\alpha}}}$ = $1.54\,\pm$\,\num{0.05E-16} \fluxunits}}\ over an aperture of diameter $1.5$\,\arcsec. This flux measurement and its associated velocity width and EW$_0$ can be compared to literature values from long-slit spectroscopy presented in \cite{Williams14}. Our flux measurement for \LAEa\ is actually somewhat smaller than the literature result ({\mbox{F$_{{\rm{Ly\alpha}}}$ = $2.53 \pm$\num{0.08e-16} \fluxunits)}}, possibly due to the halo contaminating the end of the long slit\footnote{See figure 1 in \citealt{Williams14} for their slit placement.} but we find a large velocity width of {\sc{fwhm}}\,=$1149\pm45$ \kms, almost consistent with the literature value ({\sc{fwhm}}\,=$1381\pm124$ \kms). We also measure the equivalent width, and find \LAEaEW. This is a fairly large EW$_0$, although smaller than the published estimate (EW$_0 = 103\,\pm\,15$ \AA) from long-slit spectroscopy. The difference is possibly explained by factors such as contamination of the long slit by light from the QSO (which would boost the measured Ly$\alpha$ flux and hence the EW$_0$). Even so, our flux estimate may be subject to overestimating the Ly$\alpha$ originating within \LAEa\ as we too observe the object through the QSO's Ly$\alpha$ halo, and fit a 1D Gaussian fit to this line. Conversely however, factors such as underestimating the true extent on-sky of \LAEa\ and the inaccuracy of a Gaussian fit may result in our missing some flux.

\LAEb\ has a low Ly$\alpha$ flux with emission that appears diffuse on-sky, and with a large velocity width. Taking an aperture of 1.5\,\arcsec\ in diameter driven by the object's appearance in HST imaging, we measure a flux of {\mbox{F$_{{\rm{Ly\alpha}}}$\,=\,$0.54\,\pm$\,\num{0.05e-16}\,\fluxunits}}, and a velocity width of {\sc{fwhm}}\,=\,$2480\,\pm275$\,\kms. The numbers we derive are similar to previous results (\citealt{Williams14} measure a flux of {\mbox{F$_{{\rm{Ly\alpha}}}$\,=\,$0.33\,\pm$\,\num{0.06e-16}\,\fluxunits}}, a velocity width of {\sc{fwhm}}\,=$1225\pm257$ \kms, and an EW of {\mbox{EW$_0$ = $67\pm15$ \AA)}}. \\

In the vicinity of the SMG we measure faint Ly$\alpha$ emission, but are unable to force a fit to this line in order to measure a flux, velocity width, or EW$_0$. Summing the flux over 50\,\AA\ (for consistency with \LAEa), in an aperture of diameter 1.5\,\arcsec, we measure a low-level of Ly$\alpha$ emission at {\mbox{F$_{{\rm{Ly\alpha}}}$ = $0.04\,\pm$\,\num{0.01e-16} \fluxunits.}} \\

Next we measure a Ly$\alpha$ flux and line properties for the newly discovered source \LAEc. We find a flux of {\mbox{F$_{{\rm{Ly\alpha}}}$ = $0.24\,\pm$\,\num{0.03e-16} \fluxunits}}\ over a diameter of 1.5\,\arcsec. The velocity width measures {\sc{fwhm}}\,=\,$471\pm62$ \kms, significantly narrower than \LAEa, or indeed any of the other objects in the system. We do not see continuum emission from \LAEc\ in the \MUSE\ data, therefore we place a lower limit on EW$_0$ of \LAEc. To do so, we use the RMS of an image of $\Delta\lambda = 700$\,\AA, redward of the emission line, to place an upper limit on the continuum, and in combination with the flux measurement, derive a $5\,\sigma$ upper limit of \LAEcEW.\\

The final object for which we measure a Ly$\alpha$ flux is ``\LAEd". Motivated by an alignment of dust continuum emission and a compact object seen in \HST, we extracted a spectrum at the \HST\ position to search for Ly$\alpha$ emission. We see no evidence for significant Ly$\alpha$ emission originating from the object seen in HST imaging, although a marginal detection is seen $\sim0.2\,\arcsec$ South. We report in Table \ref{tab:Lya props} the flux seen in a $1$ \arcsec\ aperture at this position, and the redshift if the corresponding tentative detection in the \ALMA\ cube is \CII\ emission.

\begin{figure*}
    \centering
    \includegraphics[width=0.95\textwidth]{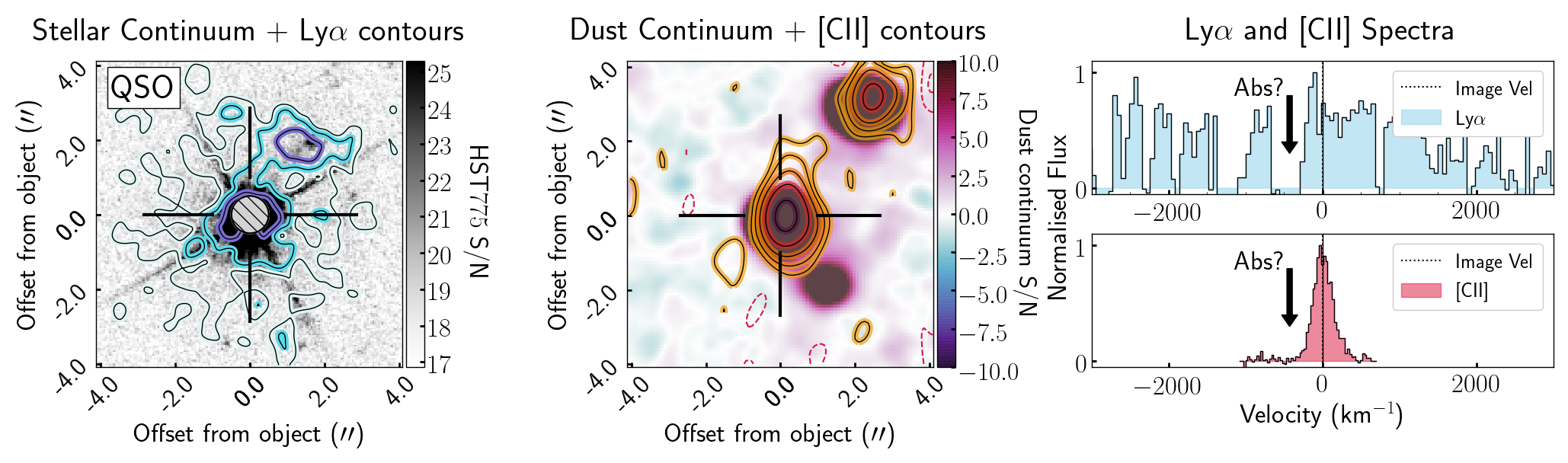}
	\includegraphics[width=0.95\textwidth]{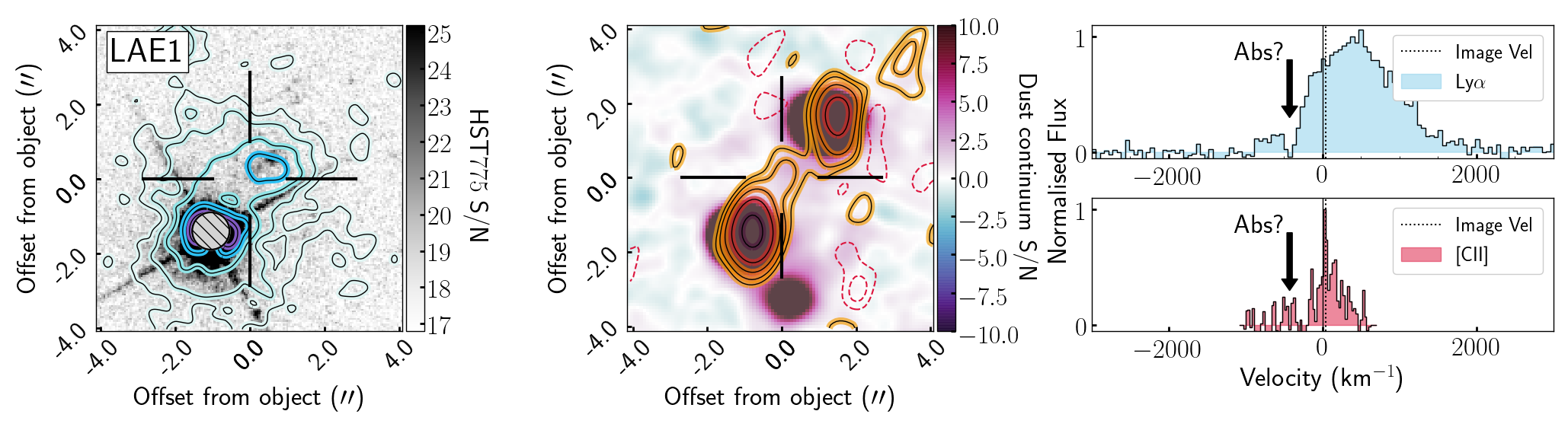}
	\includegraphics[width=0.95\textwidth]{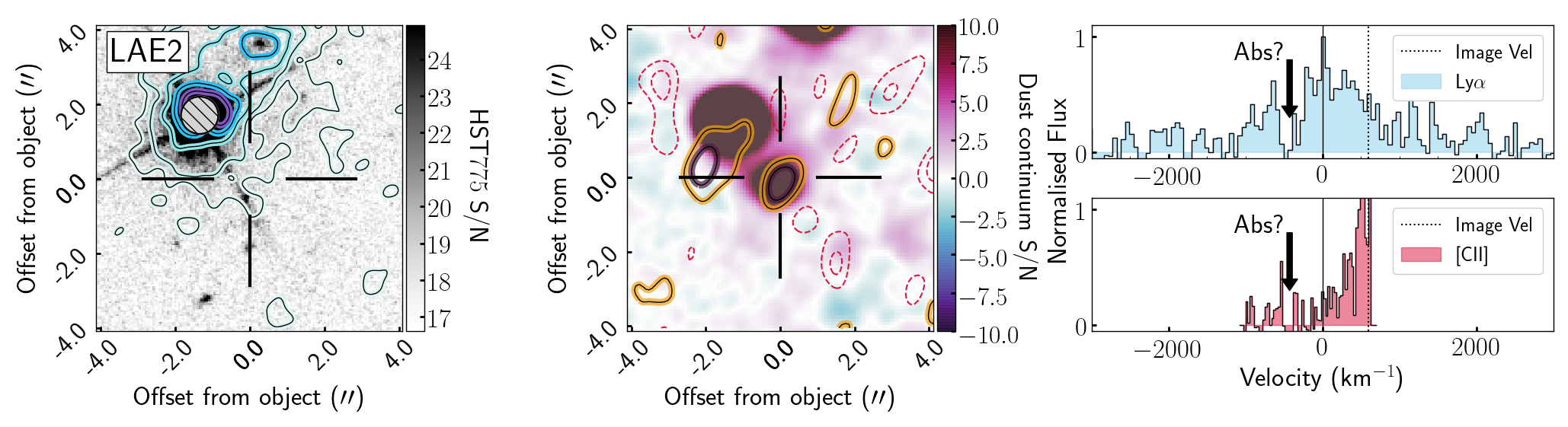}
	\includegraphics[width=0.95\textwidth]{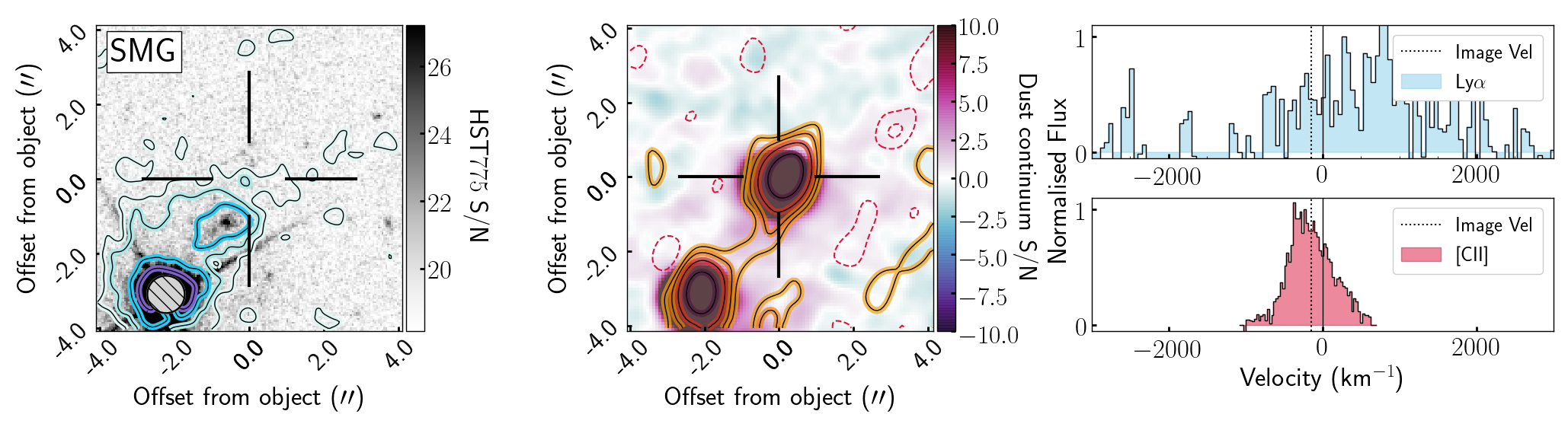}

	    \label{Fig: nature of objs}
	    	\caption{Cutout images and spectra of companion objects in the vicinity of the QSO-SMG system at $z \sim 4.7$. In the left-hand panels we show a cutout of the object in the \HST\ image, overlaid with Ly$\alpha$ contours (log-spaced with the lowest contour at 1$\sigma$) from the \MUSE\ cube, at the velocity of interest (either $\lambda$ Ly$\alpha_{{\rm{[pred]}}}$, or the peak of the Ly$\alpha$ emission if there is no \CII\ detection). In the central panels we take the same approach using the sub-mm data, and show a cutout of the object from the ALMA dust continuum map, overlaid with contoured \CII\ emission at the relevant velocity (contours are log-spaced with the lowest contour at 1.5 $\sigma$). In the right-hand panels we show Ly$\alpha$ and \CII\ spectra extracted within the aperture shown on the left-hand side. On the spectra a dotted black line represents the velocity shown in contours. For the relevant objects (The QSO, \LAEa\ and \LAEb) we incude an arrow to indicate the coherent absorption feature seen in these objects. Figure continues on next page.}
\end{figure*} 

\begin{figure*}\ContinuedFloat
	\includegraphics[width=0.95\textwidth]{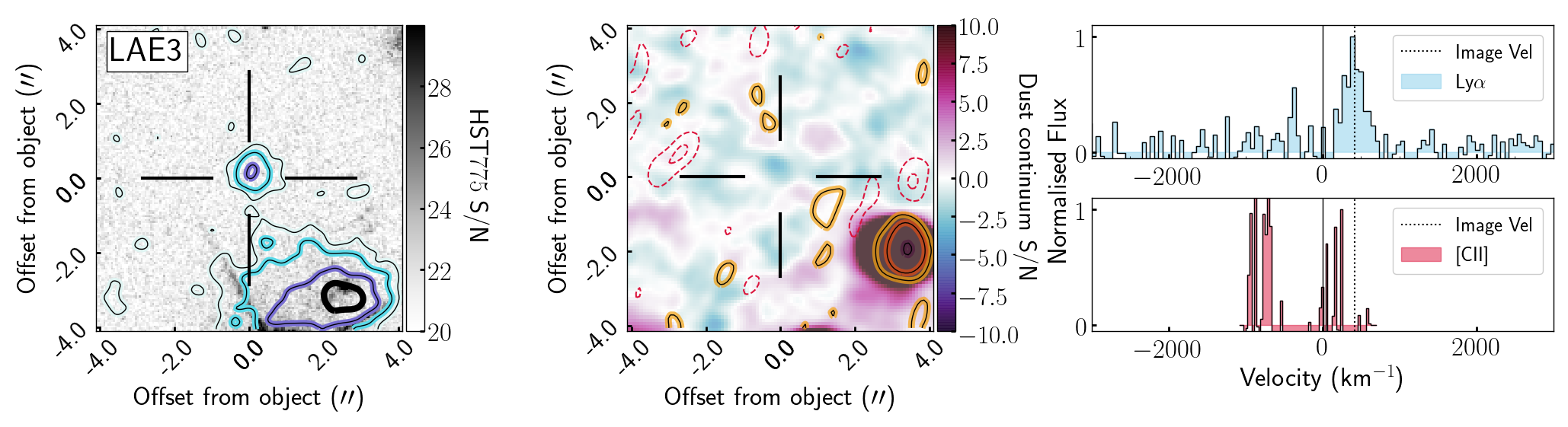}
	\includegraphics[width=0.95\textwidth]{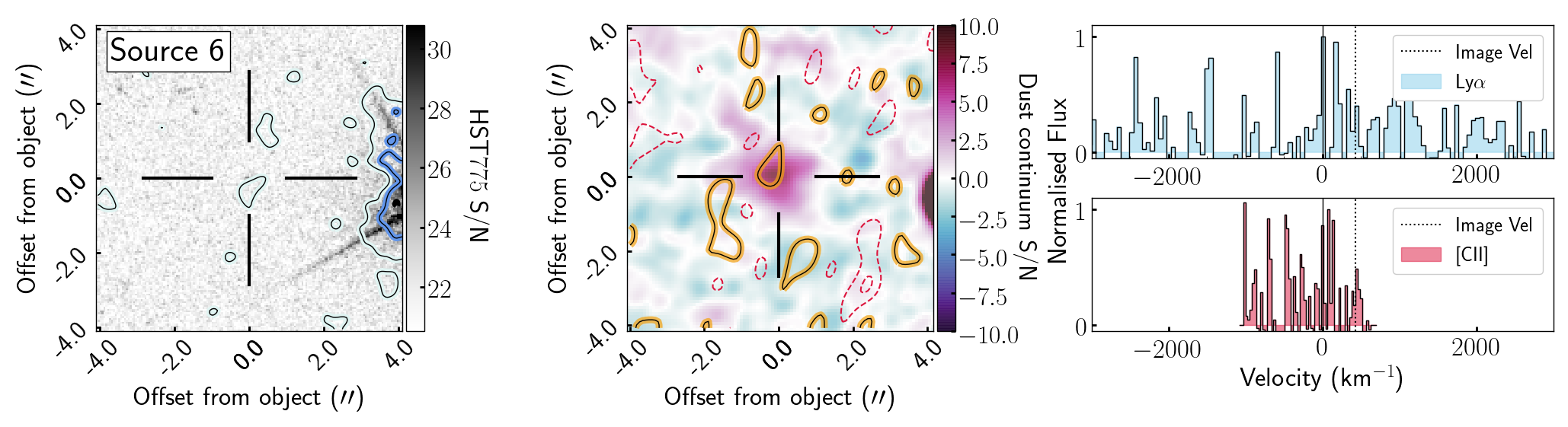}
     \caption{Continued as on previous page.}
    \label{Fig: nature of objs}
\end{figure*}

% Obj table
\begin{table*}
	\centering
	\caption{Ly$\alpha$ measurements for each object in the system. The first column gives the name for each object. The second column gives a position (see text for details). The third column gives the best estimate of the object's redshift; either the \CII\ redshift from \cite{Carilli2013}, or the Ly$\alpha$ redshift measured here. The fourth column gives the observed peak wavelength of Ly$\alpha$ emission from fitting the peak of the 1D spectra. The fifth column gives the flux within an aperture of $1.5\,\arcsec$ in diameter (except for the QSO halo, and \LAEd\ - see footnotes). The sixth column gives the velocity width of the Ly$\alpha$ line (FWHM), where it was possible to fit the 1D spectrum with a Gaussian. The seventh column gives the rest-frame Ly$\alpha$ equivalent width (or a $5\sigma$ lower limit) where appropriate. The last column gives the SFR calculated from each object's Ly$\alpha$ emission, again where appropriate.}
	
	\label{tab:Lya props}
	\begin{tabular}{lcccccccr} % columns, alignment for each
		\hline \hline
{\bf{Object}} & 

{\bf{Position}} &

{\bf{$z$$_{{\rm{best}}}$}} &

{\bf{$\lambda_{{\rm{Obs\,Ly\alpha}}}$}} & 

{\bf{F$_{{\rm{Ly\alpha}}}$} ($10^{-16}$)}  & 

{\sc{{\bf{fwhm}}}}$_{{\rm{Ly\alpha}}}$  & 
{\bf{EW$_{0\,{\rm{Ly}}\alpha}$}}\footnote{ EW$_0$ = (F$_{{\rm{Ly\alpha}}}$/F$_{{\rm{cont}}}$) $\times$ ($1$/($1$ + z$_{{\rm{best}}}$)).}  & {\bf{SFR$_{\rm{Ly\alpha}}$}}  \\

& [J2000] & & \multicolumn{1}{C}{{\rm{[\AA]}}} & 	 [\fluxunits]  & \multicolumn{1}{C}{[{\rm{km\,s$^{-1}$}}]}  & [\AA] & [M$_{\odot}$ yr$^{-1}$] \\
\hline

{\bf{QSO}} & J120523.13-074232.6 & 4.6942\footnote{\label{first} z$_{\rm{[CII]}}$} &	No peak  & \halofluxno\,\footnote{For the QSO halo we report the flux summed within the SB $>10^{-17}$ \SBunits\ contour across the entire field. Both the residuals from PSF subtraction, and the position of \LAEa\ are masked.} &	$-$  &  $-$ & $296.0\pm4.0$\\

{\bf{SMG}} & J120522.98-074229.5 & 4.6915\footref{first} & 6941.86   & 0.12$\pm$0.02 & $-$ &   $-$ & $-$ \\ 

{\bf{\LAEa}} & J120523.06-074231.2 & 4.6950\footref{first} & 6932.25  &  1.54$\pm$0.05  & 1149$\pm$45 & \LAEaEWno & 32.9$\pm$3.3 \\ 

{\bf{\LAEb}} & J120523.04-074234.3 & 4.7055\footref{first} & 6936.00\,(fix) &  0.54$\pm$0.05 & 2480$\pm$275 & $-$   & $-$ \\ 
%$\geq$310

{\bf{\LAEc}} & J120523.19-074227.7 & 4.7019\footnote{\label{second} z$_{{\rm{Lya}}}$} & 6931.65  &    0.24$\pm$0.03 &  471$\pm$62 & \LAEcEWno$_{\,5\sigma}^{\,lim}$ & 5.0$\pm$0.5 \\ 

{\bf{\LAEd}} & J120523.46-074232.1 & 4.7035\footref{first} & 6932.44\,(fix)  & 0.03$\pm$0.01 & $-$ &   $-$ & $-$ \\ 

\hline
	\end{tabular}
\end{table*}

\subsection{Nature of companions}
\label{sect: nature of objs}
% \textcolor{red}
In combination with our measurements of Ly$\alpha$ emission from the sources in this field, we use existing \ALMA\ data to compare the Ly$\alpha$ and \CII\  images and spectra, together with high-resolution optical \HST\ images, and our dust-continuum map from \ALMA. We evaluate the most likely physical scenarios leading to each object's appearance in these datasets based on Figure \ref{Fig: nature of objs}, also taking note of the channel maps shown in Appendix \ref{App:channel maps}. In each row of Figure \ref{Fig: nature of objs} we show a single object. In the left-hand panel, we show an \HST\ image representative of the optical stellar light, overlaid with Ly$\alpha$ contours extracted from the MUSE cube at the redshift of the object. These cutouts demonstrate the extended nature of Ly$\alpha$ emission compared to our previous best impression of the objects' optical sizes. In the central panel we show a dust-continuum map, overlaid with contours depicting \CII\ emission at each object's peak velocity. Finally in the right-hand panels we show the Ly$\alpha$ and \CII\ spectra in velocity space, relative to the systemic redshift of the QSO.

Interestingly, for those objects with \CII\ emission (and hence a systemic redshift), it is not evident that Ly$\alpha$ emission is always present at the same velocity, or even at the objects' positions on-sky at all. Likewise, the brightest objects shining in Ly$\alpha$ do not always correspond to a detection in \CII.

\subsubsection{QSO}
The QSO's continuum and PSF-subtracted spectrum were shown in Figure \ref{Fig: Blob Spec}, and we discussed the flux measurement in Section \ref{sect:totflux}. No distinct `peak' is seen at the predicted wavelength of Ly$\alpha$, but copious amounts of emission remain across a broad range of velocities. As seen in Figure \ref{Fig: Blob Spec}, when the emission greater than \num{1E-17} \SBunits\ is summed, the line profile becomes more centrally concentrated in wavelength. The most striking feature of this spectrum is the distinct absorption feature seen at $\approx6912$ \AA. The absorption saturates indicating a high column density of HI gas approximately $400$ \kms\ from the QSO. This kind of signature has sometimes been interpreted in the literature as an expanding shell of neutral gas surrounding a QSO (e.g. \citealt{vanojik97}, \citealt{Binette00}).

\subsubsection{SMG}

In the literature, optical emission from the SMG has evaded detection. As described above, in the MUSE datacube we place an aperture at the position of the SMG and are able to measure Ly$\alpha$ emission which would be equivalent to an instantaneous \mbox{SFR$_{\rm{Ly\alpha}}$ $\approx2$ M$_{\odot}$ yr$^{-1}$}. The line profile however is remarkably similar to the QSO's Ly$\alpha$ halo, and indeed no source is visible by eye in the Ly$\alpha$ channel maps at the position of SMG. We conclude that the Ly$\alpha$ line detected at this position arises from the edge of the QSO's extended Ly$\alpha$ halo, and so we choose not to report a Ly$\alpha$ velocity width, EW$_0$ or SFR for the SMG in Table \ref{tab:Lya props}. The channel maps in Appendix \ref{App:channel maps} present interesting features in the Ly$\alpha$ emission in the vicinity of the SMG. In particular, multiple channels spanning a few hundred \kms\ display very low surface brightness emission elongated west of the SMG.  

\subsubsection{\LAEa}
\LAEa\ is well-studied, and the measurements we report here are broadly consistent with literature results. As already noted by other studies, including \cite{Williams14}, the EW$_0$(Ly$\alpha$) is high, although it is consistent with being powered by star formation. In addition \cite{Williams14} argue that since no \CIV\ or \HeII\ emission is detected in this source, there can be at most a 10\% contribution to \LAEa's Ly$\alpha$ flux from AGN-powered photoionisation. So in conclusion there may be multiple powering mechanisms at work for the Ly$\alpha$ emission falling into the aperture. (A) Some star formation occurring within \LAEa, (B) perhaps up to a $10$\% contribution from an AGN, and (C) the Ly$\alpha$ halo extending from the QSO, within which \LAEa\ is embedded. Previous measurements of the SFR using various techniques reported values for \LAEa\ in the range \mbox{SFR$_{\rm{UV}}$$=13$\SFRunits}\ \citep{Ohyama2004}, to \mbox{SFR$_{\rm{\CII}}$$=19$\SFRunits}\ \citep{Williams14}. Transforming our Ly$\alpha$ flux to a star formation rate  we find an \mbox{SFR$_{\rm{Ly\alpha}}\approx32\pm3$ M$_{\odot}$ yr$^{-1}$}. We also note that the Ly$\alpha$ line profile is symmetric, which is not always typical of LAEs at high redshift due to the very low neutral fraction ($10^{-4}$) required to absorb all photons blue-ward of 1215.67 \AA, often leading to an asymmetric line (e.g. \citealt{Fan2001a}). \cite{Williams14} were the first to infer that the symmetric profile of \LAEa\ places a constraint on the size of the ionised bubble surrounding QSO \BR, and we concur with this conclusion. Finally we note that the absorption feature visible clearly in the spectrum of the QSO at \mbox{$\approx\,6912$\,\AA}\ is also seen in the spectrum of \LAEa, requiring that photons from the LAE as well as the QSO (some $15$\,pkpc apart) encounter the same barrier of neutral gas. 

 The peak of the Ly$\alpha$ emission from \LAEa\ falls at 6932 \AA, offset in velocity by {\mbox{$\sim$397 \kms}} from the predicted position according to the peak of \CII\ in \LAEa. Note this is greater than the previously reported value in the literature due to our velocity frame correction (Section \ref{sect:obs-ALMA cii}). Upon close inspection of the channel maps shown in Appendix \ref{App:channel maps}, it is interesting to note that the \CII\ coordinate reported for \LAEa\ (most apparent in the channel at 23.93 \kms\ is offset by $\approx 0.6$ \arcsec, although the aperture within which we extract spectra and measure flux contains both peaks. Furthermore, the purported \CII\ ``bridge" of emission between the QSO and the SMG, appears to `swirl' around the position of \LAEa -- appearing for instance both to the East and to the West of the Ly$\alpha$ peak of \LAEa\ at velocities either side of the peak. Assuming that both the \CII\ and Ly$\alpha$ emission originate from the source known as \LAEa, this could be indicative for example of tidal disruption of \LAEa's ISM during a merger between itself, the SMG, and the QSO. Regardless, we conclude that \LAEa\ is indeed likely to be a star-forming LAE associated with this group of merging galaxies, and embedded within the extended Ly$\alpha$ halo of the QSO, possibly shielded by an expanding `shell' of optically thick gas giving rise to the absorption feature.

\subsubsection{\LAEb}
 As we reported in section \ref{sect: Lya props}, the Ly$\alpha$ line that we measure is broad, and offset in velocity from the \CII\ line at the position of \LAEb. Until now this has not presented difficulty for the physical interpretation of Ly$\alpha$ emission in the vicinity of \LAEb. Taking advantage of the simultaneous spectral and spatial coverage of \MUSE\ however, we see that the Ly$\alpha$ emission line profile is remarkably similar to that from the QSO's halo. Furthermore, although \LAEb\ appears distinct in \HST\ imaging, stepping through channels of the MUSE cube, no particular layer shows spatially-peaked Ly$\alpha$ emission relative to the diffuse emission across the field. Specifically, we check the predicted wavelength of Ly$\alpha$ from \LAEb's \CII\ emission, but in addition extend the search outward to both positive and negative velocities. No spectral region suggests the presence of an object in the MUSE datacube. This implies that perhaps the emission seen in \HST\ is stellar continuum, which is outshone by diffuse Ly$\alpha$ in the \BR\ halo in the \MUSE\ cube. Indeed, we see in the channel maps that diffuse Ly$\alpha$ emission appears to extend towards the position of \LAEb\ in the velocity channels preceding \LAEb\ -- we speculate that \LAEb\ is perhaps passing through the QSO's Ly$\alpha$ halo, with a peculiar velocity directly away from the observer. From the information added here from \MUSE, we conclude that the object often referred to in the literature as `\LAEb' is in fact not responsible for powering the Ly$\alpha$ emission seen at this position in previous datasets, and this Ly$\alpha$ line is entirely consistent with being part of the extended halo surrounding QSO \BR. We do once again see the absorption feature at 6912\AA, indicating that this position on sky is also covered by the absorbing gas.

\subsubsection{\LAEc}
% \textcolor{red}
The newly discovered \LAEc\ shows a bright Ly$\alpha$ line peaking at 6931\AA, placing the LAE at a redshift of $z_{Ly\alpha} = 4.7019$. We measure properties for \LAEc\ more consistent with typical LAEs at  high redshift e.g. \cite{Drake2017a}, \cite{Drake2017b}, \cite{Hashimoto17}, \citealt{Wisotzki16}, \citealt{Leclercq2017}. The measured flux for \LAEc\ translates to an SFR = 5$\pm$0.5 \SFRunits. We place a lower limit on the EW$_0$ of \LAEc\ of \LAEcEW. This is consistent with in-situ star formation as the primary power source of the Ly$\alpha$ line, although we cannot rule out contributions from an AGN, or perhaps the edge of the QSO's halo also. 

No \CII\ or dust continuum emission is detected at the position of \LAEc. This is not altogether surprising, given the young ages and low metallicities typical of the high-redshift LAE population.  
Therefore \LAEc\ is most likely to be a star-forming LAE at the redshift of the QSO. Interestingly, the profile of the Ly$\alpha$ line appears symmetric just as in \LAEa. Following the arguments in \cite{Williams14}, we here propose that the symmetric profile of \LAEc\ indicates a new lower limit on the radius of the HII sphere/proximity zone of QSO \BR, at some $\sim 30$ pkpc away (also see \citealt{Bosman19} for a discussion on proximate LAEs at high redshift). Finally, the coherent absorption feature seen in the spectra of the other objects is at a bluer wavelength than the edge of this narrow Ly$\alpha$ line, and as such we cannot say if the proposed absorbing shell extends to the position of \LAEc.\\

\subsubsection{\LAEd}
The team\LAEd\ is detected primarily in the dust continuum. In addition, we see tentative \CII\ emission coincident with part of the area of dust emission, and an even less convincing Ly$\alpha$ detection. If the \CII\ detection is real, this places the object at $z = 4.7025$. It is notable also, that low-surface-brightness Ly$\alpha$ emission in this region could once again be part of the QSO's Ly$\alpha$ halo, appearing coincident with an unrelated object along the line of sight.

\section{Summary} \label{sect:concl}

We have presented new optical IFU data from \MUSE\ across the \BR\ field, which is one of the most overdense regions of the early Universe known at $z=4.69$. \BR\ is clearly a very extreme system, undergoing a major merger and demonstrating drastic tidal disruption of the smaller galaxies in the vicinity. Prior to this work \BR\ was already one of the most studied single-targets at high redshift, acting as a laboratory for the study of diverse galaxies typically detected via different selection techniques, all evolving in the same environment. Here, we examine the Ly$\alpha$ halo surrounding QSO \BR\ and measure Ly$\alpha$ properties for companions in the field, including a new LAE discovered in this work. In conjunction with existing \ALMA\ observations we examine and compare the neutral and ionised gas content of the \BR\ system's Ly$\alpha$ halo and constituent objects, which provides us with the best view to date of the physical processes under way in \BR\ until new facilities (e.g. JWST and the GTO programme) become available. Our main findings can be summarised as follows:

\begin{itemize}
    \item QSO \BR\ exhibits a large Ly$\alpha$ halo, stretching across at least $\sim$ 8\,\arcsec\ on-sky ($\approx 55$ pkpc at $z=4.6942$) at surface brightness levels greater than \num{1E-17} \SBunits.
    
    \item In contrast, no Ly$\alpha$ halo is detected around the SMG, which has a similar gas mass and SFR as the QSO. 
    
    \item We do not find evidence for extended \CIV\ emission surrounding the QSO down to a surface brightness limit of SB\,$\approx$\,\num{0.19E-16}\,\SBunits\ in a square arcsecond, and hence can not place any constraint on the dominant powering mechanism of the Ly$\alpha$ halo.
    
    \item The optical NW-companion of QSO \BR\ known as ``\LAEa" appears to be a Ly$\alpha$-emitting galaxy embedded within the larger region of diffuse Ly$\alpha$ emission. Taking an aperture of 2\,\arcsec\ in diameter we measure a Ly$\alpha$ flux, velocity width consistent with literature measurements from long-slit spectra. We measure an \LAEaEW, this is probably due to a combination of powering mechanisms for the Ly$\alpha$.
    
    \item The optical SW-companion of QSO \BR, confirmed in \CII\ and \NII\ emission, known as ``\LAEb" is not necessarily responsible for the Ly$\alpha$ emission previously reported at this position. Although we detect a Ly$\alpha$ line, it appears diffuse on-sky, with no local maximum coinciding with the position of \LAEb. The Ly$\alpha$ line profile, peak, and FWHM are also more consistent with being part of the QSO's diffuse halo.
    
    \item We detect an additional LAE in the \BR\ system $\sim5$\,\arcsec\ to the north of the QSO, and denote it ``\LAEc". The object was discovered serendipitously in the \MUSE\ datacube, and exhibits a bright but narrow Ly$\alpha$ line of flux $0.24$\,$\pm$ $\num{0.03E-16}$ \fluxunits\,(1.5\,\arcsec), a velocity width of 471.23 \kms\ and \LAEcEW. This LAE is more aligned with typical properties of Ly$\alpha$ emitters at high redshift, with Ly$\alpha$ emission consistent with powering by star formation. 
    
    \item The symmetric profile of Ly$\alpha$ in \LAEc\ places a new constraint on the size of the HII bubble surrounding QSO \BR\ of $\approx\,60$\,pkpc in diameter.
    
    \item We see coherent absorption, possibly indicative of an expanding shell of HI gas, 400 \kms\ infront of \BR\ across at least 4\,\arcsec\ on-sky, corresponding to $\sim$24 pkpc (diameter). The feature is blue-ward of the edge of Ly$\alpha$ from \LAEc\ and hence we cannot confirm whether \LAEc\ is covered by the HI gas.
\end{itemize}

These results demonstrate the efficiency of \MUSE\ for detecting low-surface-brightness Ly$\alpha$ emission in addition to the instrument's use as a `redshift machine' for the detection of emission-line galaxies. Futhermore, the combined power of \MUSE\ and \ALMA\ to examine both ionised- and cool gas components offers unprecedented insights on the nature of a diverse population of objects at high redshift. This work also  highlights the need for follow-up of Ly$\alpha$ halos with  future facilities (such as JWST) with a view to unambiguously detecting/ruling out the presence of e.g. metal lines, to finally disentangle the processes powering these giant structures.

\acknowledgments

ABD, FW, MN and MN acknowledge support from the ERC Advanced Grant 740246 (Cosmic Gas). \\

This paper makes use of the following ALMA data: 
ADS/JAO.ALMA\#2011.0.00006.SV, 2013.1.00039.S, \break 
2013.1.00259.S, 2013.1.00745.S, 2015.1.00388.S, \break
2015.1.01489.S, 2017.1.00963.S, 2017.1.01516.S.
ALMA is a partnership of ESO (representing its member states), NSF (USA) and NINS (Japan), together with NRC (Canada), NSC and ASIAA (Taiwan), and KASI (Republic of Korea), in cooperation with the Republic of Chile. The Joint ALMA Observatory is operated by ESO, AUI/NRAO and NAOJ. \\

\software{Astropy \citep{2013A&A...558A..33A}; CASA  \citep{mcmullin07}, MUSE data reduction pipeline \citep{Weilbacher12}, \citep{Weilbacher14}, ZAP \citep{Soto2016}, MPDAF \citep{Piqueras17}.}

\appendix
\section{Appendix A}

\label{App: flux measurements}
We include here in Figure \ref{Fig: Fluxes} the spectra and images from which we measure the Ly$\alpha$ properties of each source. In the left-hand panels we show a cutout of the object in the high-resolution \HST\ image, in the central panels we show a spectrum extracted from the \MUSE\ cube within the aperture shown on the \HST\ image, and in the right-hand panels we show a cutout image extracted from the \MUSE\ cube across the wavelength range shaded in orange on the spectrum. Apertures are centred on the \CII\ positions reported in \cite{Carilli2013} where possible, otherwise the peak of the Ly$\alpha$ emission seen in \MUSE\ is taken. For the objects where no 1D Gaussian fit could be performed, we shade in red the region of the spectrum which is summed to estimate the Ly$\alpha$ flux.

\begin{figure*}
    \centering
    \includegraphics[width=0.95\textwidth]{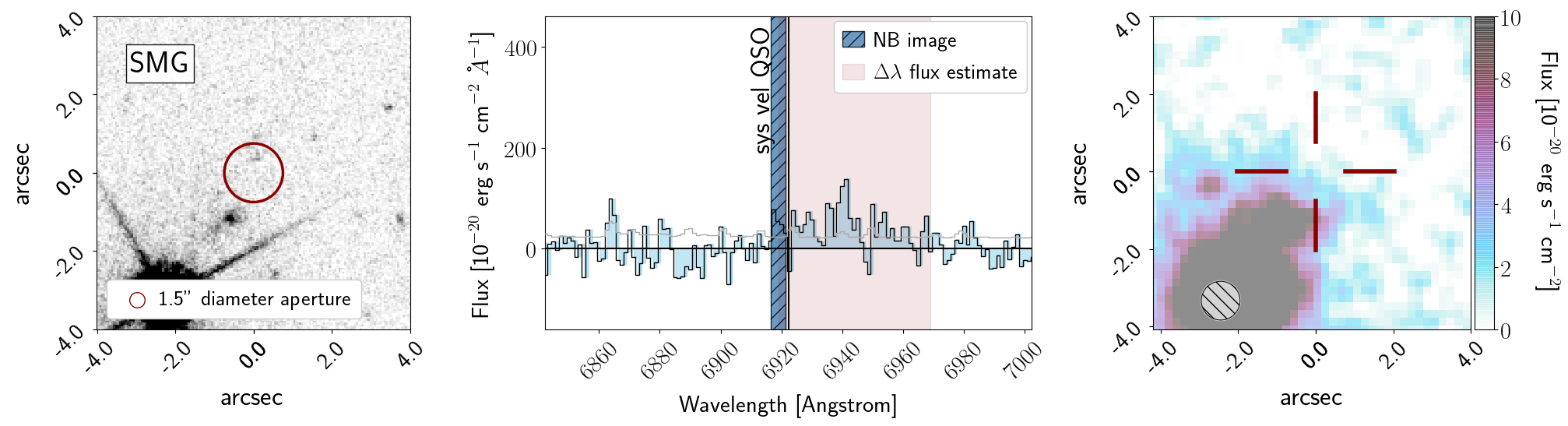}
	\includegraphics[width=0.95\textwidth]{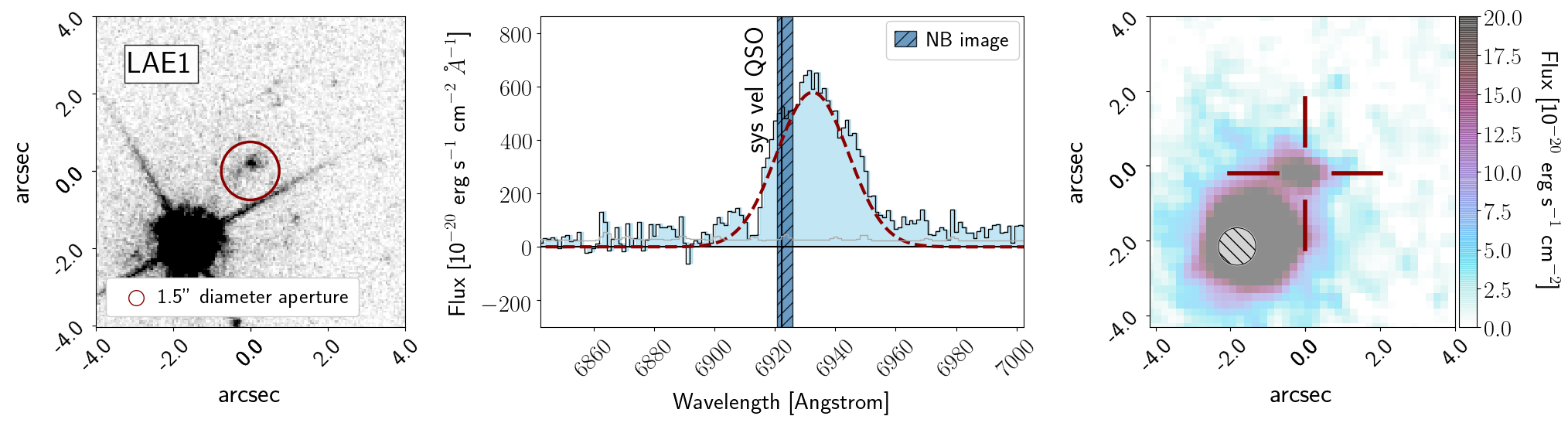}
	\includegraphics[width=0.95\textwidth]{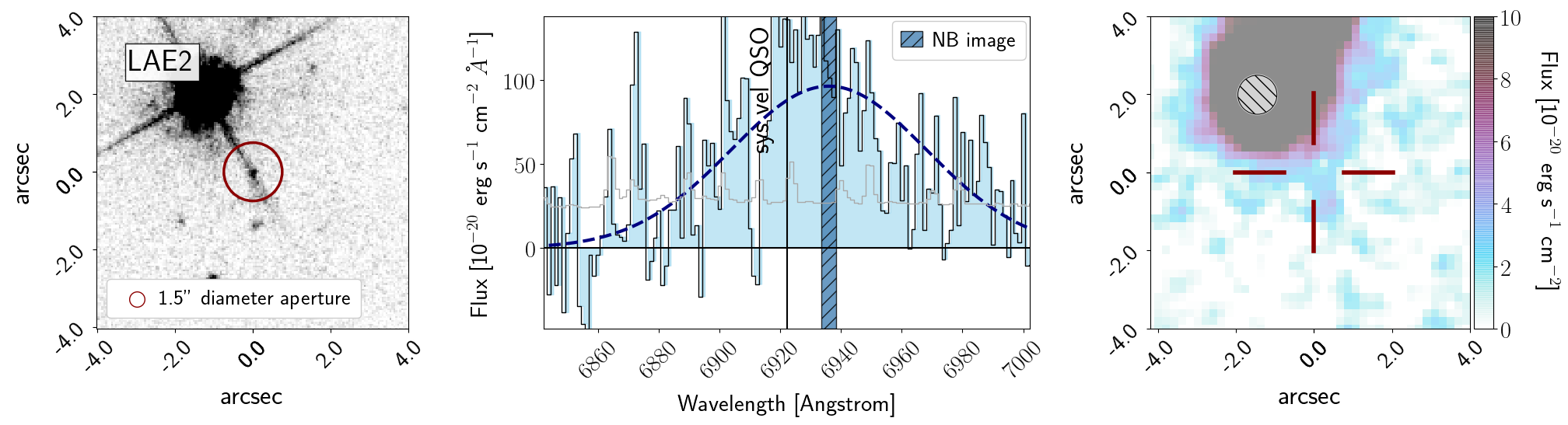}
	\caption{Cutout images and spectra of companion objects in the vicinity of the QSO-SMG system at $z \sim 4.7$. Figure continues on next page.}
\end{figure*} 

\begin{figure*}\ContinuedFloat
	\includegraphics[width=0.95\textwidth]{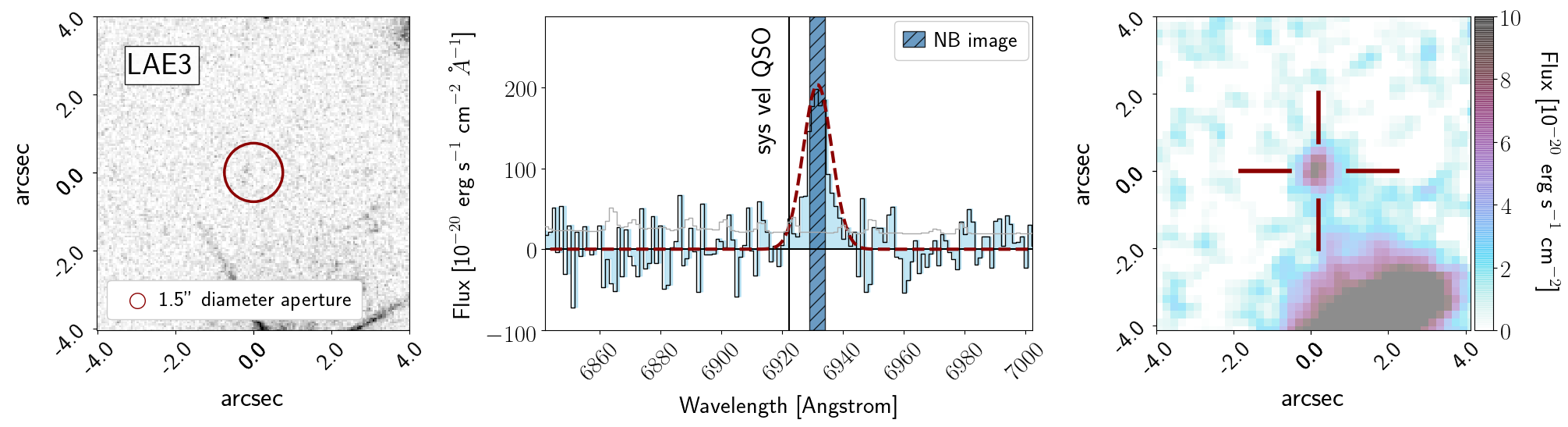}	\includegraphics[width=0.95\textwidth]{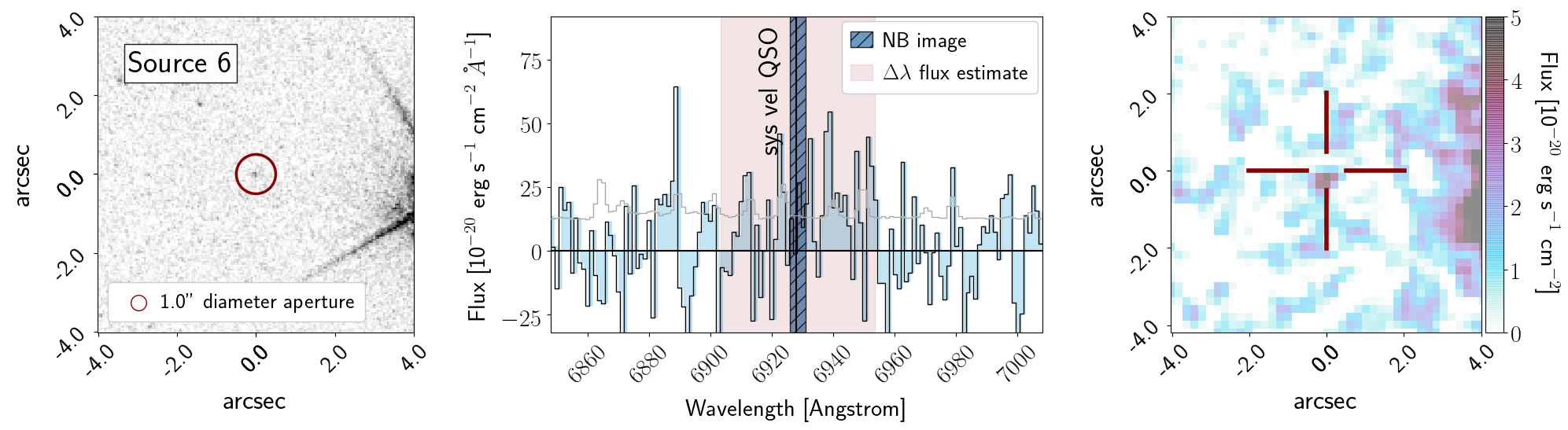}
	\caption{Continued as on previous page.}

    \label{Fig: Fluxes}
\end{figure*}

\section{Appendix B}
\label{App:channel maps}
This appendix contains the \MUSE\ and \ALMA\ channel maps depicting the extended Ly$\alpha$ and \CII\ emission at a series of velocities relative to the QSO's systemic redshift in Figure \ref{Fig: channel maps}. 

\begin{figure*}
    \centering
	\includegraphics[width=0.99\textwidth]{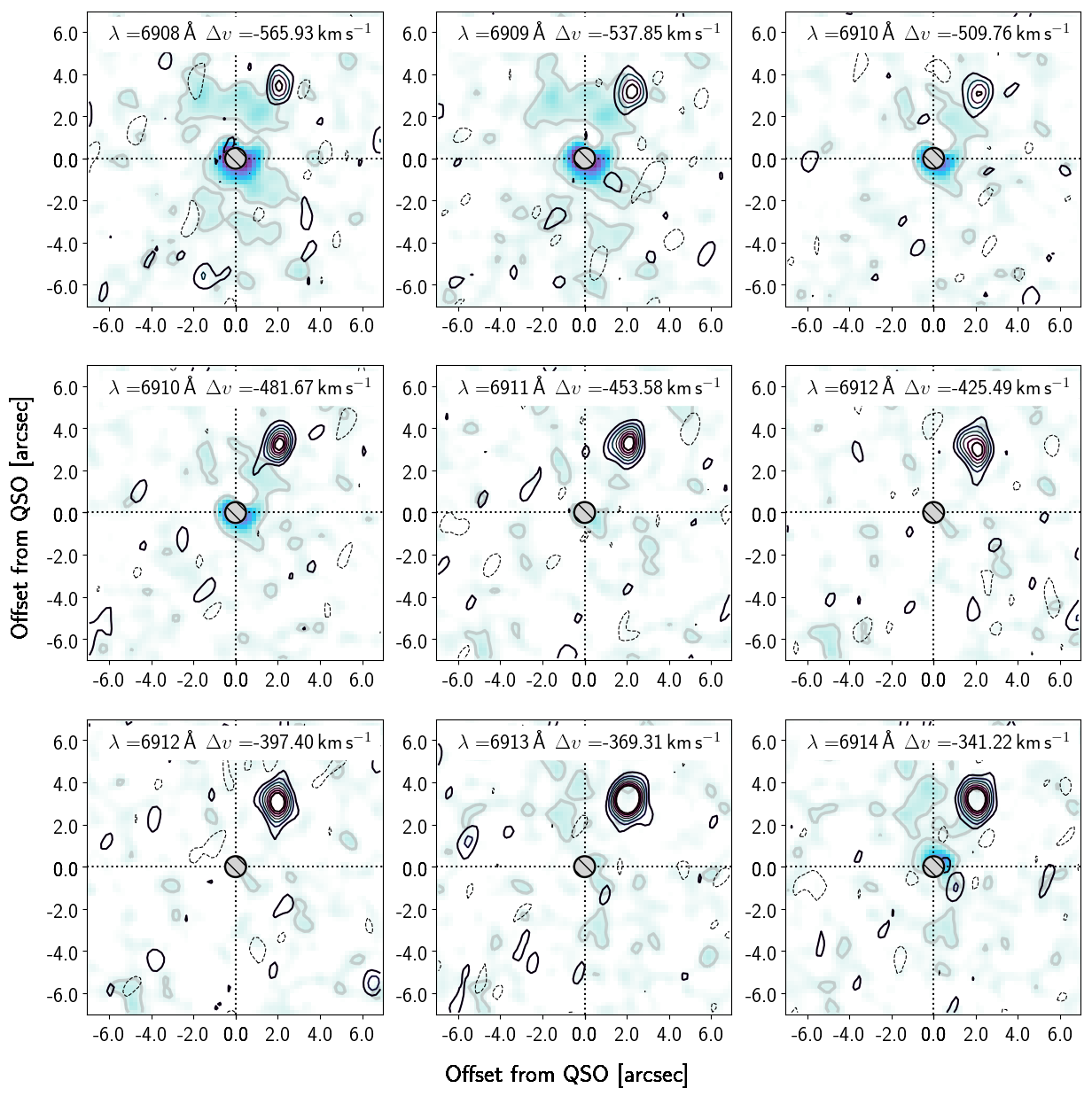}
  \caption{\MUSE\ Ly$\alpha$ channel-maps, overlaid with \CII\ contours from \ALMA, where contours are linearly spaced at $\pm$ $1.5, 3.0, 6.0, 9.0, 12.0$ and $15.0\,\sigma$, and negative contours are represented by dotted lines. \MUSE\ data have been rebinned in velocity for comparison to the lower velocity resolution of the \ALMA\ channels. Figure continues on next page.}
  \end{figure*} 

\begin{figure*}\ContinuedFloat
    \centering
	\includegraphics[width=0.99\textwidth]{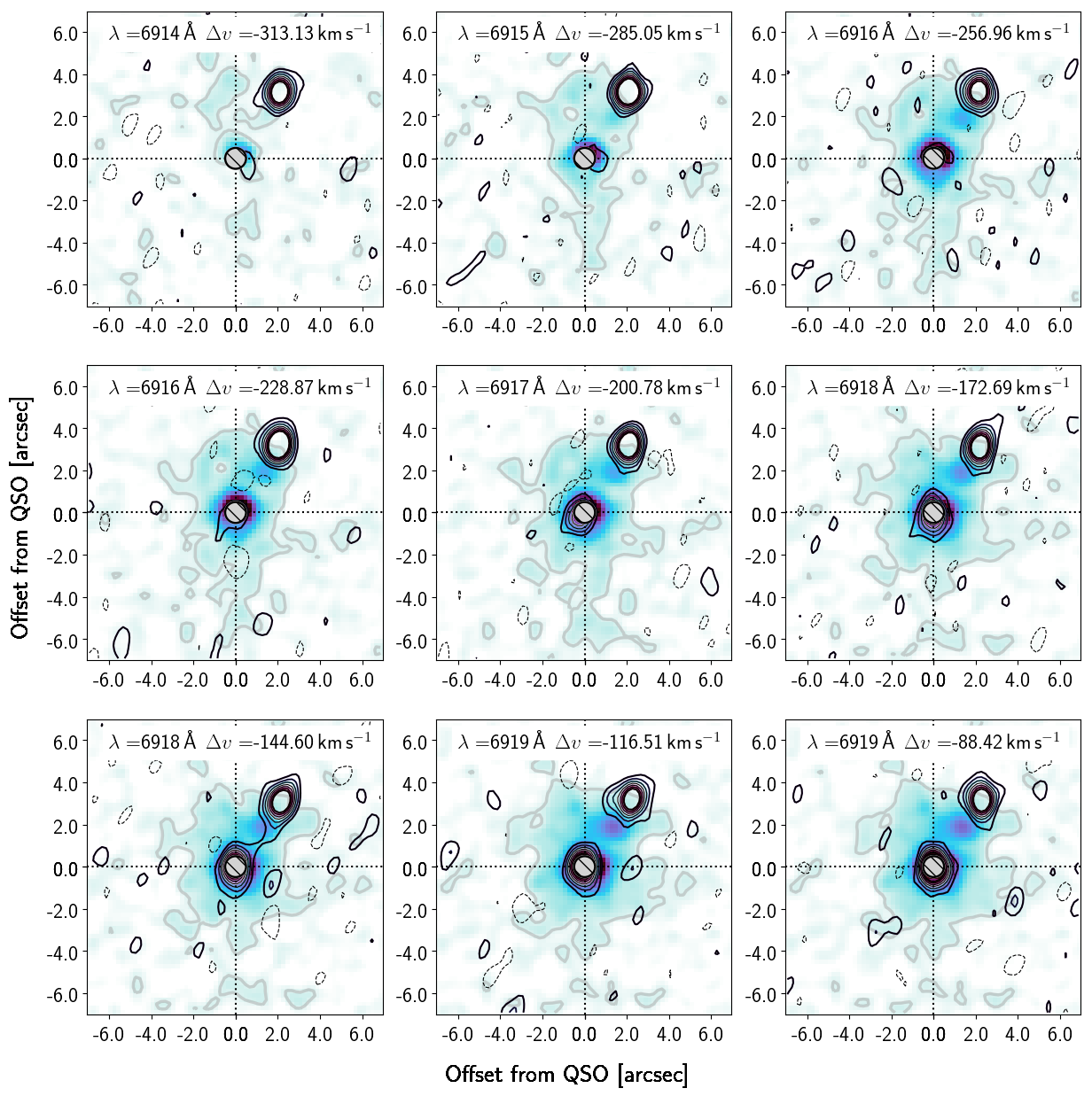}
	\caption{Continued from previous page.}
\end{figure*} 

\begin{figure*}\ContinuedFloat
    \centering
	\includegraphics[width=0.99\textwidth]{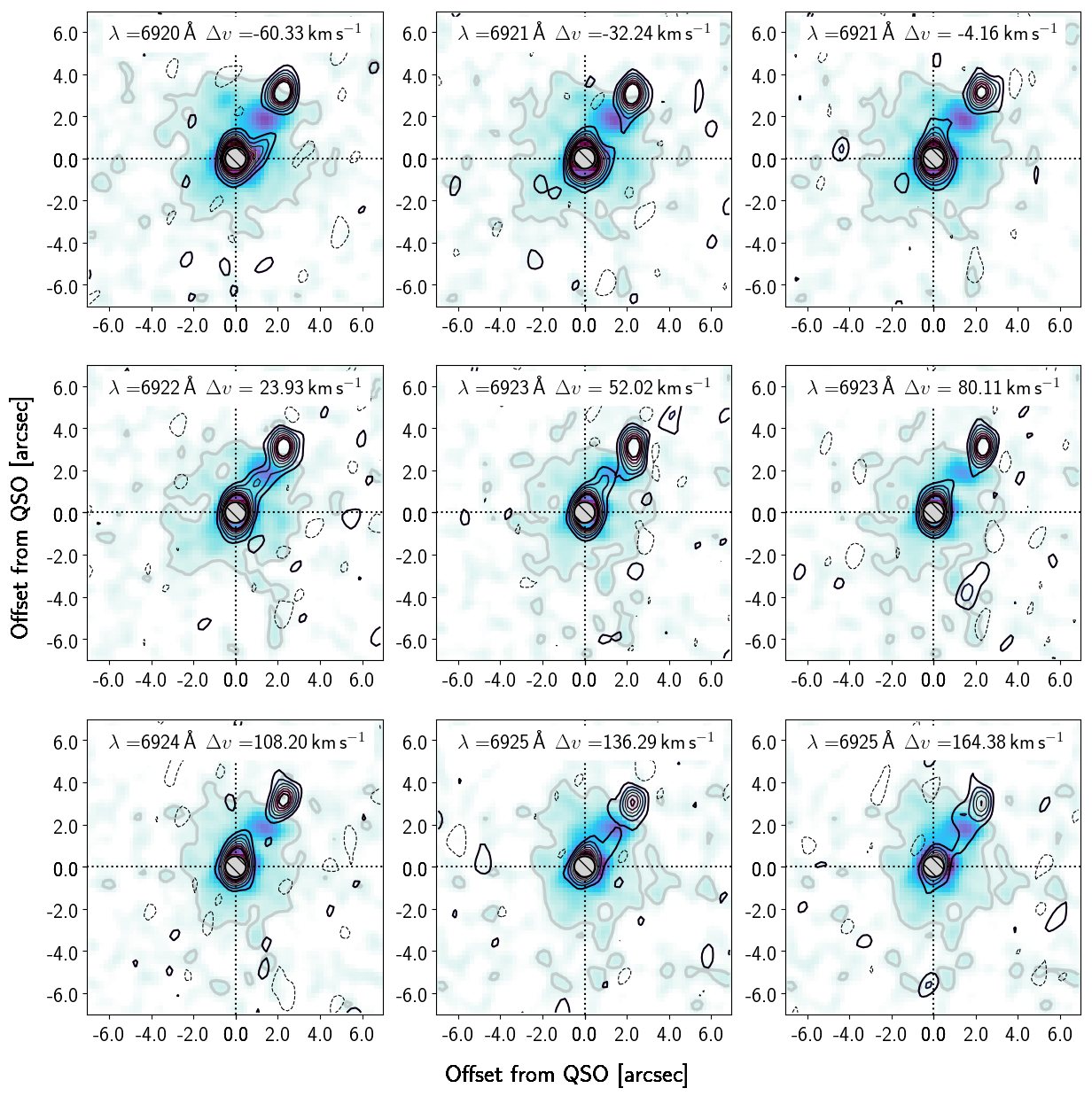}
	\caption{Continued from previous page.}
\end{figure*} 

\begin{figure*}\ContinuedFloat
    \centering
	\includegraphics[width=0.99\textwidth]{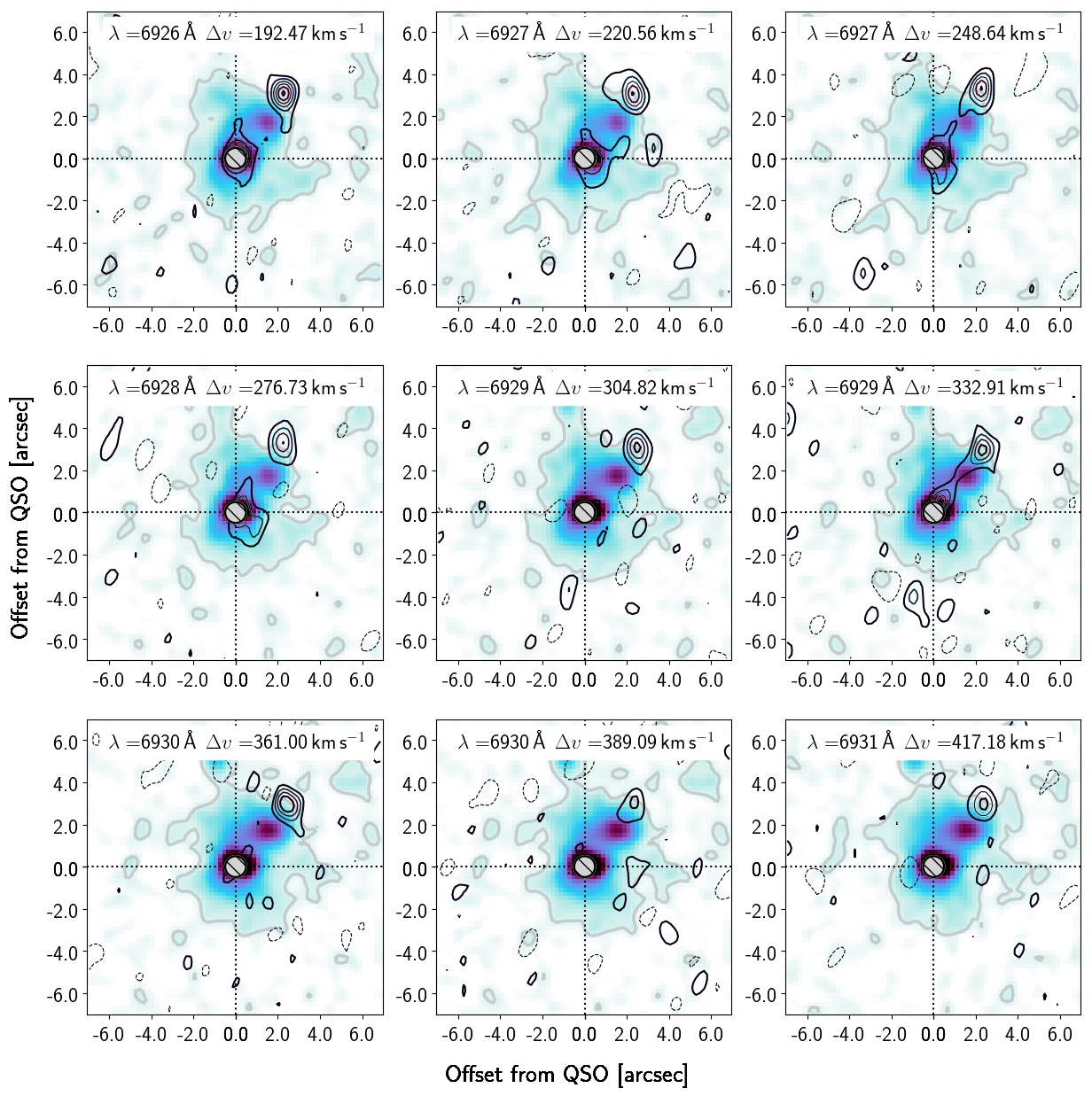}
	\caption{Continued from previous page.}
\end{figure*}

\begin{figure*}\ContinuedFloat
    \centering
	\includegraphics[width=0.99\textwidth]{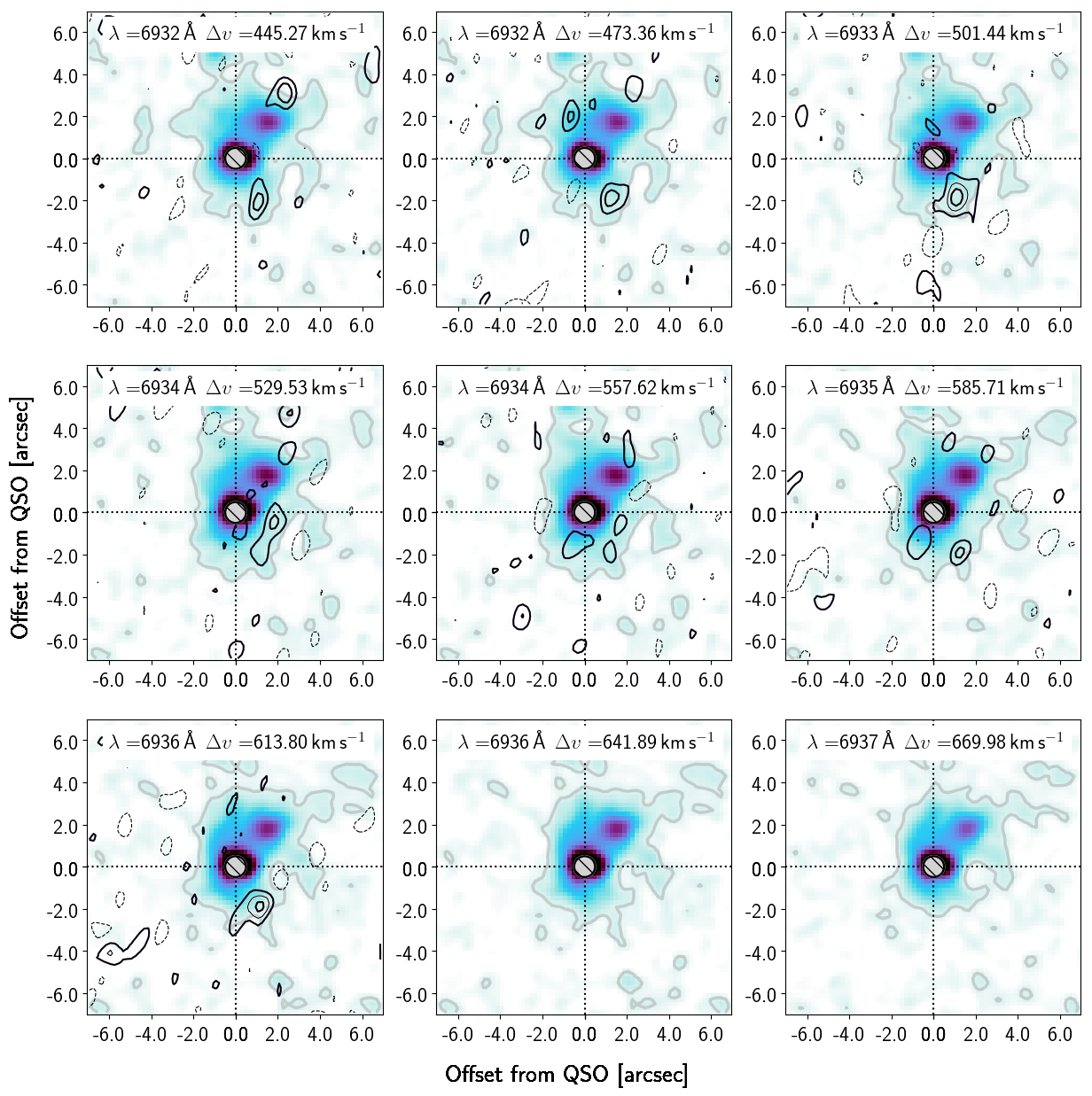}
    \label{Fig: channel maps}
    \caption{Continued from previous page.}

 \end{figure*} 

\section{Appendix C}
\label{APP: CIV}
This appendix demonstrates our search for extended \CIV\ emission surrounding QSO \BR\ discussed in Section \ref{sect:powering}.
\begin{figure*}
    \centering
    \includegraphics[width=0.95\textwidth]{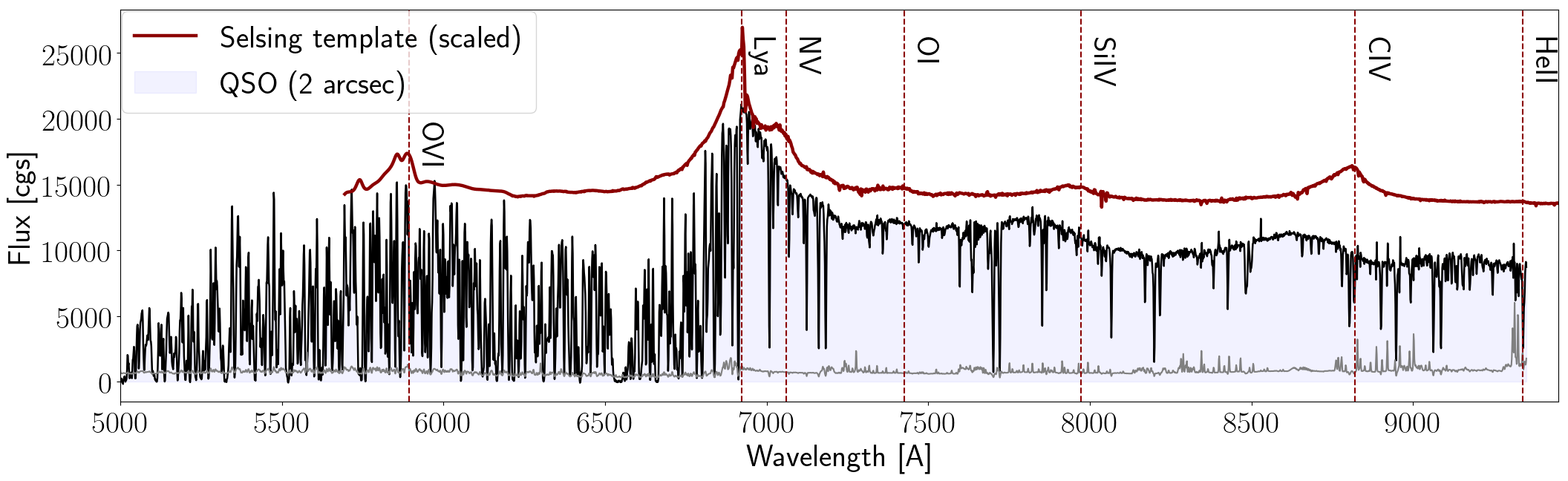}
    \includegraphics[width=0.75\textwidth]{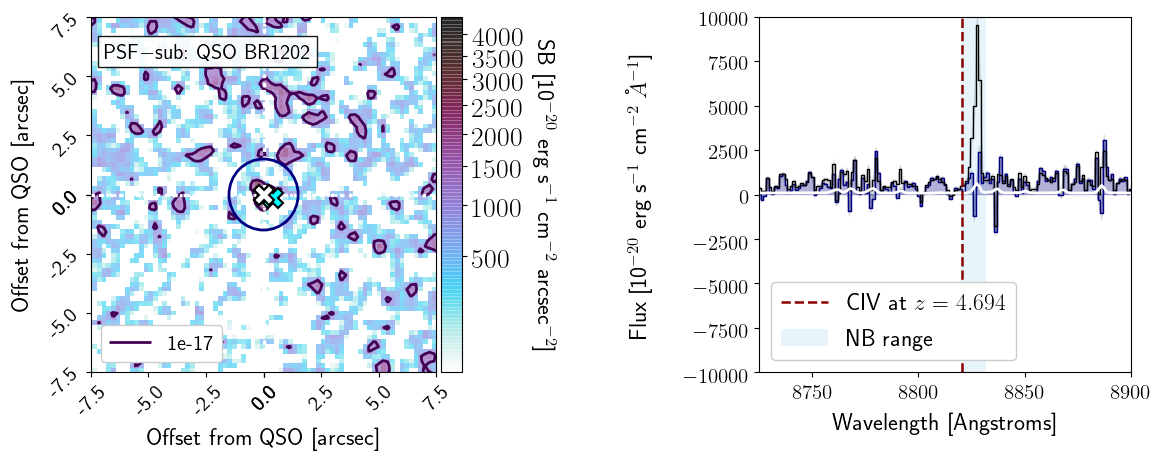}
    \caption{In the top panel we show the QSO spectrum extracted in a two arcsecond diameter aperture (black line, filled blue) and its associated noise (grey). This is overlaid with a QSO template from \cite{Selsing16} (dark red, offset for clarity) which gives the predicted positions of emission lines in the spectrum. In the lower two panels we show the same two cutouts as in Figure \ref{Fig: Blob Spec}, however this time centred on the blue-shifted wavelength region corresponding to \CIV\ emission. No extended \CIV\ emission is detected. }
    \label{fig:my_label}
\end{figure*}

%\begin{thebibliography}{}

%\bibitem[Astropy Collaboration et al.(2013)]{2013A&A...558A..33A} Astropy Collaboration, Robitaille, T.~P., Tollerud, E.~J., et al.\ 2013, \aap, 558, A33 
%\end{thebibliography}

%\bibliography{bib} % if your bibtex file is called example.bib

\end{document}